\documentclass[twocolumn,times]{aastex63}

\defcitealias{Nitschai2020}{\scshape Paper I}

\received{January 26, 2021}
\revised{May 19, 2021}
\accepted{May 20, 2021}

\submitjournal{ApJ}

\shorttitle{Dynamical Model of the Milky Way}
\shortauthors{}

\begin{document}

\title{Dynamical model of the Milky Way using APOGEE and \textit{Gaia} data}

\correspondingauthor{Maria Selina Nitschai}
\email{nitschai@mpia.de}
\author[0000-0002-2941-4480]{Maria Selina Nitschai}
\affiliation{Max Planck Institute for Astronomy, K{\"o}nigstuhl 17, 69117 Heidelberg, Germany}

\author[0000-0003-2895-6218]{Anna-Christina Eilers}\thanks{NASA Hubble Fellow}
\affiliation{MIT Kavli Institute for Astrophysics and Space Research, 77 Massachusetts Ave., Cambridge, MA 02139, USA}

\author[0000-0002-6922-2598]{Nadine Neumayer}
\affiliation{Max Planck Institute for Astronomy, K{\"o}nigstuhl 17, 69117 Heidelberg, Germany}

\author[0000-0002-1283-8420]{Michele Cappellari}
\affiliation{Sub-department of Astrophysics, Department of Physics, University of Oxford, Denys Wilkinson Building, Keble Road, Oxford OX1 3RH, UK}

\author[0000-0003-4996-9069]{Hans-Walter Rix}
\affiliation{Max Planck Institute for Astronomy, K{\"o}nigstuhl 17, 69117 Heidelberg, Germany}

\begin{abstract}
We construct a dynamical model of the Milky Way disk from a data set, which combines \textit{Gaia} EDR3 and APOGEE data throughout Galactocentric radii between $5.0\leq R\leq19.5$~kpc. We make use of the spherically-aligned Jeans Anisotropic Method to model the stellar velocities and their velocity dispersions. Building upon our previous work, our model now is fitted to kinematic maps that have been extended to larger Galactocentric radii due to the expansion of our data set, probing the outer regions of the Galactic disk. Our best-fitting dynamical model suggests a logarithmic density slope of $\alpha_{\rm DM}=-1.602\pm0.079_{\rm syst}$ for the dark matter halo and a dark matter density of $\rho_{\rm DM}(R_{\odot})=(8.92\pm0.56_{\rm syst})\times 10^{-3}$~M$_{\odot}$~pc$^{-3}$ ($0.339\pm0.022_{\rm syst}$~GeV~cm$^{3}$). We estimate a circular velocity at the solar radius of $v_{\rm circ}=(234.7\pm1.7_{\rm syst})$~km~s$^{-1}$ with a decline towards larger radii. The total mass density is $\rho_{\rm tot}(R_{\odot})$=$(0.0672\pm0.0015_{\rm syst})$~M$_{\odot}$~pc$^{-3}$ with a slope of $\alpha_{\rm tot}$=$-2.367 \pm 0.047_{\rm syst}$ for $5\leq R\leq19.5$~kpc and the total surface density is $\Sigma(R{_\odot}, |z|\leq$ 1.1~kpc)=$(55.5\pm1.7_{\rm syst})$~M$_{\odot}$~pc$^{-2}$. While the statistical errors are small, the error budget of the derived quantities is dominated by the 3 to 7 times larger systematic uncertainties. These values are consistent with our previous determination, but systematic uncertainties are reduced due to the extended data set covering a larger spatial extent of the Milky Way disk. Furthermore, we test the influence of non-axisymmetric features on our resulting model and analyze how a flaring disk model would change our findings.
\end{abstract}


\section{Introduction} \label{sec:intro}
Dynamical models are important for our understanding of galaxies. They describe the distribution of stars over orbits in a gravitational potential \citep{Binney1987}. Hence, we can describe a galaxy as stars orbiting in a smooth gravitational potential by interpreting the combined stellar position and velocity information \citep[e.g.][]{Rix1997, Syer1996, Binney2011, JAM}. This allows us to infer the gravitational potential, the circular velocity curve, mass distribution and dark matter density of a galaxy \citep[e.g.][]{rix13}.

For most external galaxies dynamical models usually suffer from degeneracies due to limited data. Since only line-of-sight observations are available, the data cannot provide three dimensional information that is needed to fully constrain dynamical models. There are only very few cases were the full velocity information is available and usually they have a limited quality of data \citep[e.g.][]{vanDeVen2006, Watkins2015}. However, our own Galaxy can be observed in great detail with high precision measurements from various stellar surveys. In the past years there has been a plethora of surveys for the Milky Way, such as for example APOGEE \citep{APOGEE08, majewski17} and \textit{Gaia} \citep{Gaia_mission_16}, providing astrometric and spectroscopic information for a large sample of stars.

Many different dynamical models can be found in the literature \citep[see][]{rix13}, such as Jeans modelling \citep[e.g.][]{JAM, Bovy2012, Zhang2013}, action based models \citep[e.g.][]{Binney2011, Binney2012, Bovy13}, Schwarzschild models \citep[e.g.][]{Rix1997, vanderMarel1998, Gebhardt2000, Cappellari2006, vandenBosch2008} and made-to-measure models \citep[e.g.][]{Syer1996, deLorenzi2009, Dehnen2009, Portail17, Wegg15}. For this paper we will use the Jeans modelling approach. Jeans models are based on the Jeans equations \citep{Jeans1915, jeans1922}, which are derived from the steady-state Boltzmann equation under the assumption of axisymmetry. The steady-state collisionless Boltzmann equation needs to be satisfied by the distribution function of the system, which describes the position and velocity of the stars in equilibrium and steady state under the gravitational influence of a smooth potential \citep{Binney1987}. The solutions of the Jeans equations describe the moments of the velocity distribution, and the density of a collection of stars in a gravitational potential. However, to obtain a unique solution one has to assume the shape and orientation of the velocity ellipsoid. \citet{JAM} reviews the possible natural choices for this alignment, which are prolate spheroidal coordinates, spherical coordinates, and cylindrical ones.

Previously, we used the \textit{Gaia} DR2 kinematics to construct an axisymmetric dynamical model of the Milky Way disk \citep[][hereafter Paper I]{Nitschai2020}. There we used the new spherically-aligned Jeans Anisotropic Modelling \citep[JAM$_{\rm sph}$,][]{Cappellari2019} method, since the \textit{Gaia} data \citep{Hagen2019, Everall2019} showed that the velocity ellipsoid is closer to being spherically aligned than cylindrically \citep[JAM$_{\rm cyl}$,][]{JAM}. But we also compared the results against JAM$_{\rm cyl}$ and found negligible differences.

The known deviations from equilibrium and axisymmetry for the Milky Way \citep{widrow12, Antoja2018, Gaia2018_nonaxi} which are in conflict with our model assumptions, are not uncommon for other galaxies too, but models are still able to recover the average kinematic properties \citep{cappellari13}, even from less precise data. In the case of the JAM method, it has been used to model integral-field stellar kinematics of large numbers of external galaxies \citep{cappellari16_rev}. It has been tested against N-body simulations \citep{Lablanche12, Li16} and in real galaxies against CO circular velocities \citep{leung18}, including barred and non-perfectly-axisymmetric galaxies similar to the Milky Way. In both cases, it recovers unbiased density
profiles, even more accurately than the Schwarzschild \citep{Schwarzschild79} approach \citep{leung18}, which was also confirmed \citep{Jin2019} by using Illustris cosmological N-body simulation \citep{Vogelsberger2014}. Hence, we expect that JAM will be able to capture the main kinematic features of the Milky Way and give accurate results for the total density and circular velocity, even more since we use the spherical alignment, JAM$_{\rm sph}$.

In this paper we build upon \citetalias{Nitschai2020}, by extending the maximal Galactocentric radius of the data from $\sim 12$~kpc to $\sim 20$~kpc. This allows us to extend our model to the outer parts of the Galactic disk and to better constrain the model parameters. Furthermore, we improve our model by including the uncertainties of the kinematic maps at each position.
The extended data set also allows us to test our model when using a flared outer disk, as suggested by e.g. \citet{Gyuk1999_fl, Momany2006, Lopez_Corredoira2014, Li2019}.

The outline of this paper is as follows: we describe the data set and the derived kinematic maps with their uncertainties in Section~\ref{sec:data}. In Section~\ref{sec:Method} we describe the components of our mass model for the Galaxy and our modelling method. In the end, we show our results for this extended data set and the investigation in the effect of the non-axisymmetries in Section~\ref{sec:res}, and we also show the results for a model with a flared disk. Finally, in Section~\ref{sec:con} we summarize our results. In the Appendix we include further tests for our model.

\section{Data} \label{sec:data}

In this paper we combine two data sets, one are giant stars from \textit{Gaia} early data release 3 (EDR3) with radial velocities \citep{GaiaEDR3_2021} and one are the red giant branch (RGB) stars from APOGEE and \textit{Gaia} \citep{Hogg2019}. This combined data set will allow us to reach out to large Galactocentric radii.

\subsection{Data set}

\begin{figure*}[t]
\centering
 \includegraphics[width=2.1\columnwidth]{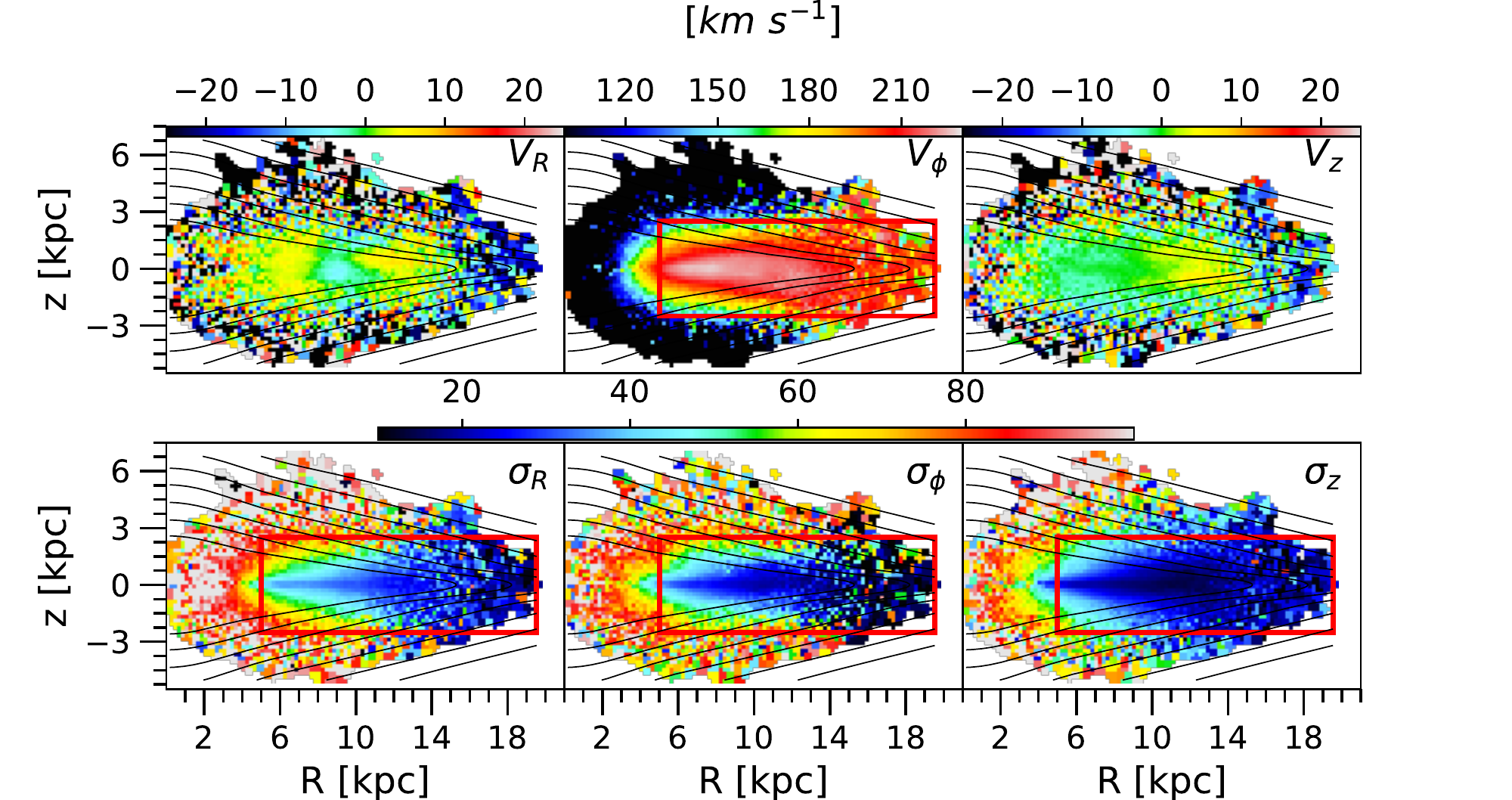}
 \caption{\textbf{Velocity maps for the combined data from \textit{Gaia} EDR3 with that from \citet{Hogg2019}}. The color represent the mean velocity or dispersion of the Voronoi bins, as described in Section~\ref{subsec: vel maps}. In the top panels are the velocities and in the bottom panels the velocity dispersions. From left to right are the velocities and velocity dispersions in $R$, $\phi$ and $z$. All color-maps are in km~s$^{-1}$. Inside the red boxes are the data that we will actually use for the JAM model and the black lines are the stellar density (as described in Section~\ref{subsec: MW main}) contours.}
\label{im:vel map}
\end{figure*}

\begin{figure*}[t]
\centering
 \includegraphics[width=2.1\columnwidth]{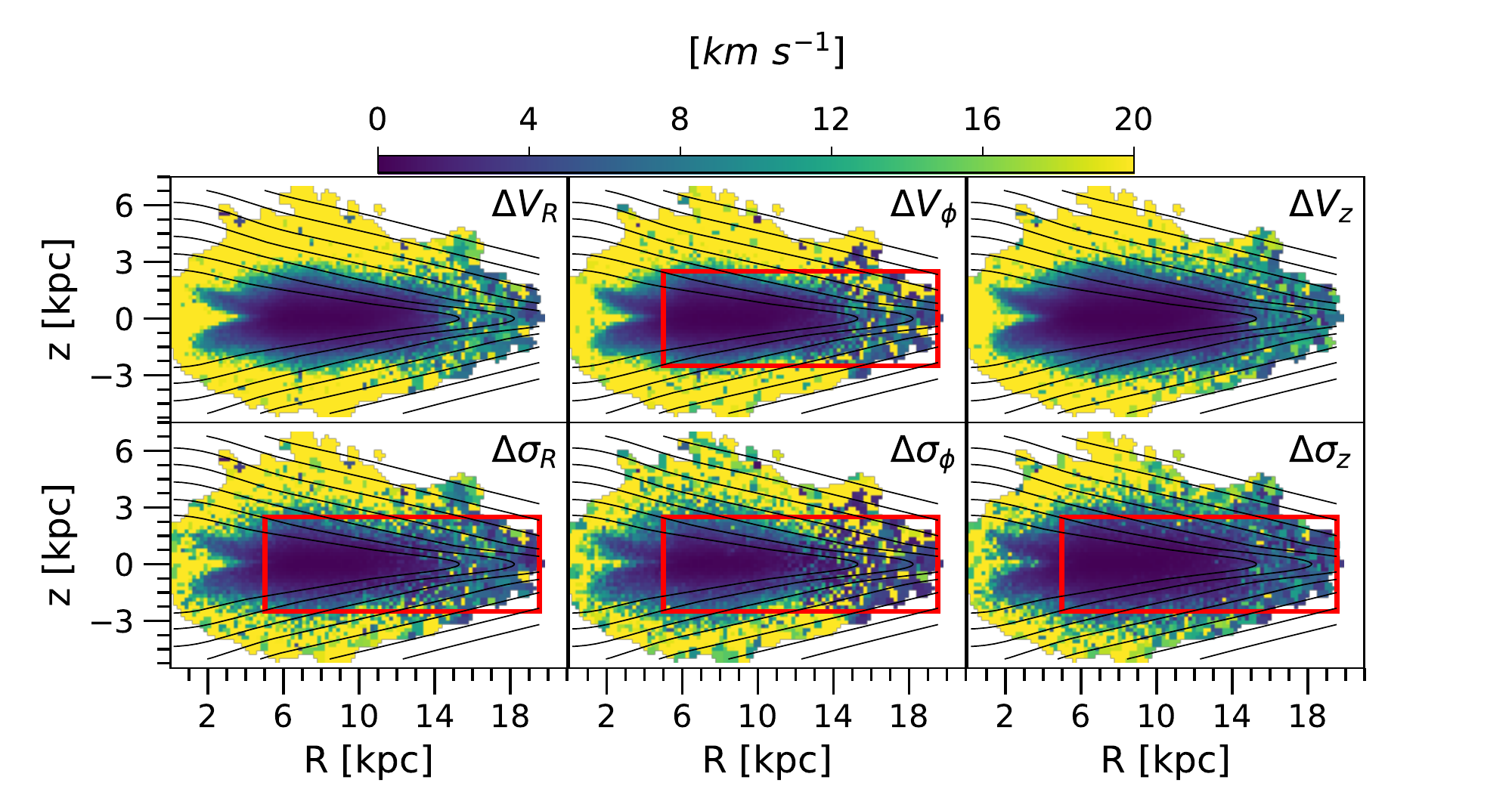}
 \caption{\textbf{Uncertainty maps for the combined data.} Similar to Figure~\ref{im:vel map}, the colors represent the uncertainty value for each Voronoi bin. In the top panels are again the velocities and in the bottom panels the velocity dispersions. From left to right are the velocities and velocity dispersions in $R$, $\phi$ and $z$. Inside the red boxes are the errors that we will use in calculating the $\chi^2$ of our JAM model and the black lines are the stellar density (Section~\ref{subsec: MW main}) contours.}
\label{im:error}
\end{figure*}

\textit{Gaia} provides precise parallax information locally but further away from the Sun the distance uncertainties dominate. In addition, it is known that there is an parallax offset in the \textit{Gaia} Data Release 2 (DR2) \citep{Lindegren2018, Zinn2019, Schoenrich19} and also in \textit{Gaia} EDR3 a parallax bias has been discovered \citep{Lindegren_2021_astrometric_sol}. There has been an attempt to map the dependencies of the parallax bias, but since the dependencies are non-trivial it is not possible to have a definitive correction \citep{Lindegren2021}.

In this work we update the data from \citetalias{Nitschai2020} using \textit{Gaia} EDR3 \citep{GaiaEDR3_2021}. The important updates for our study is the higher precision in the astrometry in EDR3, however the radial velocities have not been updated and have been adopted from DR2. In addition, EDR3 does not have extinction or reddening values, hence we cross matched EDR3 with DR2 using the given crossmatch catalog by \citet{Torra_crossmatch_2021} in order to get these values. For distances we use the photogeometric distance estimates by \citet{Bailer-Jones_2021} for the \textit{Gaia} EDR3 in order to avoid the parallax bias and high uncertainties further away from the Sun.

The selection criteria we apply are the same as we already had in \citetalias{Nitschai2020}. Since giant stars are the main contribution in \textit{Gaia} at distances larger than 1~kpc from the Sun and can be measured out to large distances due to their brightness, we select only them in this subsample. They are selected based on their absolute magnitude, $M_G  = m_G -5\cdot \log{d}+5-A < 3.9,$ given the extinction by \textit{Gaia} DR2 and d being the distance from \citet{Bailer-Jones_2021}, their intrinsic colour $(G_{BP}-G_{RP})_0 > 0.95$, with the reddening taken from \textit{Gaia} DR2 and parallaxes with relative uncertainty $\varpi/\epsilon_{\varpi} > 5$.

These data are precise only around the Sun but not far beyond.

In order to probe larger Galactocentric distances \citet{Hogg2019} obtained precise spectrophotometric distance estimates for $44,784$ red giant stars with a data-driven model. They combined spectroscopic data from APOGEE \citep{majewski17} DR14 \citep{Abolfathi2018} with photometric information from \textit{Gaia} \citep{Gaia_mission_16} DR2 \citep{Gaia2_18}, 2MASS \citep{Skrutskie2006} and WISE \citep{Wright2010}. These data allow us to probe larger distances, out to Galactic radii of $\sim 20$~kpc, than with \textit{Gaia} data alone.

These RGB stars where selected with a cut at $\log{g}<2.2$, in order to have stars more luminous than the red clump. For more details on quality cuts for these stars see \citet{Hogg2019}.

In order to extend our data set to larger Galactocentric radii, we combine the data set based on \textit{Gaia} EDR3 with that from \citet{Hogg2019}. For the combined data set we require that the proper motion and line-of-sight velocities of all stars are finite and that they are in the area of $-30^\circ<\phi<30^\circ$. For sources that are in both data sets we prefer the values from \citet{Hogg2019} for our further calculations.

The remaining stars from the \cite{Hogg2019} data are $34,180$ and $2,823,509$ stars that are only in \textit{Gaia}. In total we have $2,857,689$ stars that will be used in the further analysis.

\subsection{Kinematic maps}\label{subsec: vel maps}

In order to calculate the kinematic maps of the Milky Way, we transform the positions and velocities of the stars in our combined data set, into a Galactocentric reference frame.

We assume as distance to the Galactic centre $R_{\odot}=8.178$~kpc \citep{Abuter19}, a vertical displacement of the Sun from the midplane of $z_{\odot}=0.02$~kpc \citep{Joshi07} and as solar velocities in cylindrical Galactic coordinates $(U_{\odot}, V_{\odot}, W_{\odot})= (-11.1, 247.4, 7.2)$ km~s$^{-1}$ \citep[from][respectively]{Schoenrich10,Abuter19,Reid2009}. We require the data to have $|z| \leq$ 7~kpc because we have more stars towards positive $z$ values, which are not covering the whole area and creating gaps in our data. Hence, to not have a data set that is highly asymmetric and not continues, we exclude from the beginning stars with too high $z$ values. These stars that are excluded would anyways not be considered in our further analysis, since they are halo stars and lay outside of the disk region we want to model. We then divide our data in 200~pc $\times$ 200~pc cells and do Voronoi binning, using the \textsc{vorbin} package\footnote{We use the Python version 3.1.4 of the \textsc{vorbin} package available from \url{https://pypi.org/project/vorbin/}\label{foot:vorbin}} \citep{Cappellari2003_vorbin} on this cells, requiring a minimum of $5$ stars for each Voronoi bin. We use the Voronoi binning on our already divided into 200~pc $\times$ 200~pc data, to fully make use of the data, especially towards larger radii and $z$ where the stars are not as dense and we would not get more than $5$ stars in a cell. With that method we can combine the data from cells with not enough stars in them, but still keep as smallest bin the 200~pc $\times$ 200~pc cell in the areas where we have numerous stars and not have too tinny bins. For each bin we calculate the mean and standard deviation of the velocities. The final number of bin is $2,899$ and covers a volume with extreme cylindrical coordinates 0.18~kpc $\leq R\leq$ 19.5~kpc and -5.02~kpc $\leq z\leq$ 6.8~kpc.

The created velocity maps of the combined data set in cylindrical coordinates are shown in Figure~\ref{im:vel map}. As expected the stars around the solar position extend to larger Galactocentric heights $z$ than at larger or smaller radii. However, thanks to the Voronoi binning we get a continuous covered area up to a radius of 19.5~kpc without any gaps in the kinematics.

It is worth mentioning that the selection function of \textit{Gaia} DR2 \citep{Boubert2020_a, Boubert2020_b} and specifically the RV sample \citep{Rybizki2020}, since we only select stars in EDR3 that have radial velocities in DR2 and for APOGEE \citep{Bovy2014, Mackereth2020} will not affect our model significantly, since we study the kinematics of the observed stars, at a given position. The stellar tracer density is assumed, not fitted and the selection function at a given position does not depend strongly on the velocities of the objects.

To find the uncertainties for our kinematic maps, we choose to use the bootstrapping method \citep[e.g.][]{Efron1993}. We selected this method and not error propagation because it provides a more robust estimate of the uncertainties. The propagated uncertainties from radial velocities and proper motions measurements tend to underestimate the true uncertainties. In detail, we draw 100 random samples of stars with their velocities and redo the Voronoi binning, calculating again for each of these samples the mean velocity and the velocity dispersion for each bin. Afterwards we find the uncertainties of the bin by calculating the standard deviation of the 100 mean velocities and velocity dispersions.

The uncertainty maps are shown in Figure~\ref{im:error}. Errors around the Sun and the midplane are much smaller ($\leq$ 4~km~s$^{-1}$) than at larger heliocentric distances, because there are more stars detected and their measurements are more precise since they are closer to us.

The uncertainties are also higher towards the Galactic Center. In this area, starting at 5~kpc, the bar dominates the kinematics \citep{Wegg15, Bovy19} and hence would break the axisymmetry assumptions of our model. To avoid that, we exclude data with $R<5$~kpc. Additionally, we also exclude stars with $|z|\geq 2.5$~kpc, which is the same range we had in \citetalias{Nitschai2020}, in order to exclude halo stars. Similar to \citetalias{Nitschai2020}, we ignore the halo distribution since it is not well known and in this regime we expect to mainly have disk stars. The red boxes in Figure~\ref{im:vel map} and ~\ref{im:error} show the area we will investigate and it has a $1,404$ bins, the data outside the boxes is excluded from the further analysis.

\section{Methods}\label{sec:Method}

\subsection{Mass distribution of the Milky Way model} \label{sec:MW model}

To construct a dynamical model we need an estimate of the tracer distribution and the mass density of the Galaxy.
We will assume the same model for the Milky Way as in \citetalias{Nitschai2020}, a detailed description and explanation can be found there and we give a brief summary in this section. In addition to this model we also investigate a model with a flaring disk in Section~\ref{subsec: MW flared}.

\subsubsection{Distribution with an exponential disk model} \label{subsec: MW main}

Our main stellar model makes use of the \citet{juric08} disk stellar distribution, normalized to the local luminosity density from \citet{Flynn06}. However, we ignore the stellar halo component because in the area we are probing with our data set, the stellar mass distribution is dominated by the disk component.

The disk is decomposed into a sum of two exponential components, the thin and the thick disk with different scale lengths ($L_{\rm thin}, L_{\rm thick}$) = (2.6, 3.6)~kpc and heights ($H_{\rm thin}, H_{\rm thick}$) = (0.3, 0.9)~kpc \citep{juric08}:
\begin{equation}
    \rho_{\rm D}(R, z, L, H) = \rho(R_{\odot}, 0) \exp\left(-\frac{R-R_{\odot}}{L}- \frac{|z|}{H}\right)
\end{equation}
\begin{eqnarray}
    \rho_{\rm D}(R, z) & = & \rho_{\rm D}(R, z, L_{\rm thin}, H_{\rm thin})\nonumber\\ 
    &&+ f\cdot\rho_{\rm D}(R, z, L_{\rm thick}, H_{\rm thick})
\end{eqnarray}
where $f = 0.12$ is the thick disk normalization relative to the thin disk \citep{juric08}. In addition, we combined that with an axisymmetric bulge \citep{mcmillan17, Bissantz2002}. The luminosity density profile of this model is shown in the top panel of Figure~\ref{im:mW dens}.

We approximate this stellar model with a multi-Gaussian expansion \citep[MGE,][]{Emsellem94,Mge2002}, using the MGE fitting method \citep{Mge2002} and \textsc{mgefit} software package\footnote{We use the Python version 5.0.12 of the \textsc{mgefit} package available from \url{https://pypi.org/project/mgefit/}\label{foot:mge}}, which is necessary in order to apply the JAM model \citep{JAM, Cappellari2019}.

For the mass density of the Galaxy we need to include also the gas and dark matter contribution. For the gas component we add the H$_2$ and H$_{\textsc{ii}}$ distribution from \citet{mcmillan17}. For simplicity we ignore the hole in the center of the density profile, since the hole is outside of the range we are probing. In \citetalias{Nitschai2020} we have shown that this does not change the dynamical model significantly. Like for the stellar model, we use the \textsc{mgefit} package \citep{Mge2002} to fit the created image of the gas density. This MGE will be added to the potential density and be kept fixed during the model fit, to the quoted mass by \citep{mcmillan17}.

As a dark matter halo we assume a generalized Navarro, Frenk and White profile \citep[gNFW,][]{Wyithe2001}:
\begin{equation}
    \rho_{\rm DM} = \rho_s \left(\frac{m}{rs}\right)^{\alpha_{\rm DM}}\left(\frac{1}{2} + \frac{1}{2}\frac{m}{rs}\right)^{-3-\alpha_{\rm DM}}
\end{equation}
where:
\begin{equation}
    m^2 = R^2 + (z/q_{\rm DM})^2,
\end{equation}
$r_s$ the scale radius, $\alpha_{\rm DM}$ the dark matter slope and $q_{\rm DM}$ the axial ratio. If $\alpha_{\rm DM}$ = -1, the profile represents the classical Navaro, Frenk, and White (NFW) dark matter profile \citep{NFW96}.

The dark matter profile is a one-dimensional profile and we fit this with Gaussians using the \textsc{mge\_fit\_1d} routine from the \textsc{mgefit} package. It can be made oblate or prolate, for use with the model depending on the value of $q_{\rm DM}$.

The model assumptions cause systematic uncertainties that strongly depend on the values we choose for the different parameters but also from the model itself. We have tested the effects of the disk parameters in \citetalias{Nitschai2020} for the circular velocity and total density, in order to get an understanding of these uncertainties.

\subsubsection{Distribution with a flared disk model} \label{subsec: MW flared}

We know that the outer part of our Galaxy, beyond 15~kpc, is warped and flared, but the details of its shape are still uncertain \citet{review16}. These structures were first detected in the gas component of the disk \citep{Kerr1957, Oort1958, Grabelsky1987, May1997}, but can also be detected in the three dimensional distribution of stars \citep[e.g.][]{Liu2017,Anders2019, Skowron2019}.

A flared disk means that the vertical scale height grows with increasing radius. The stellar flare in the outer Galactic disk has been observed and the scale height modeled as a function of the radius by many studies for the outer disk \citep[e.g.][]{Gyuk1999_fl, Alard2000, Lopez_Corredoira2002, Yusifov2004, Momany2006, Reyle2009, Lopez_Corredoira2014, Li2019}.

Since, we are probing a large range of Galactic radii up to $\sim 20$~kpc, our data extends to the regime where the flare is visible. Hence, we investigate, in addition to our main exponential model described in the previous subsection, how a flared disk would change our results.

Even though, there are many studies of the flare there is not one parametrization for it. We decided to base the flare we use on the work of \citep{Lopez_Corredoira2014}, because they give a parametrization for the scale height for the thin and thick disk. In detail, we keep the density profile the same as for our main model and only modify the scale heights, $H_{\rm thin}, H_{\rm thick}$, for larger radii:

\begin{equation}
    H_{\rm thin} = H_{\rm thin}(R_{\odot})\left[1 + \sum \limits_{i=1}^{2} k_{i, \rm thin} \left(R-R_{\odot}\right)^i\right], R > R_{\odot}
\end{equation}
\begin{equation}
    H_{\rm thick} = H_{\rm thick}(R_{\rm ft})\left[1 + \sum \limits_{i=1}^{2} k_{i, \rm thick} \left(R-R_{\rm ft}\right)^i\right], R > R_{\rm ft}
\end{equation}
with $k_{i, \rm thin} = \rm{(-0.037~kpc^{-1}, 0.052~kpc^{-2})}$, $k_{i, \rm thick} = \rm{(0.021~kpc^{-1}, 0.006~kpc^{-2})}$ and R$_{\rm ft}$ = 6.9~kpc \citep{Lopez_Corredoira2014}.

\begin{figure}
\centering
\includegraphics[width=1\columnwidth]{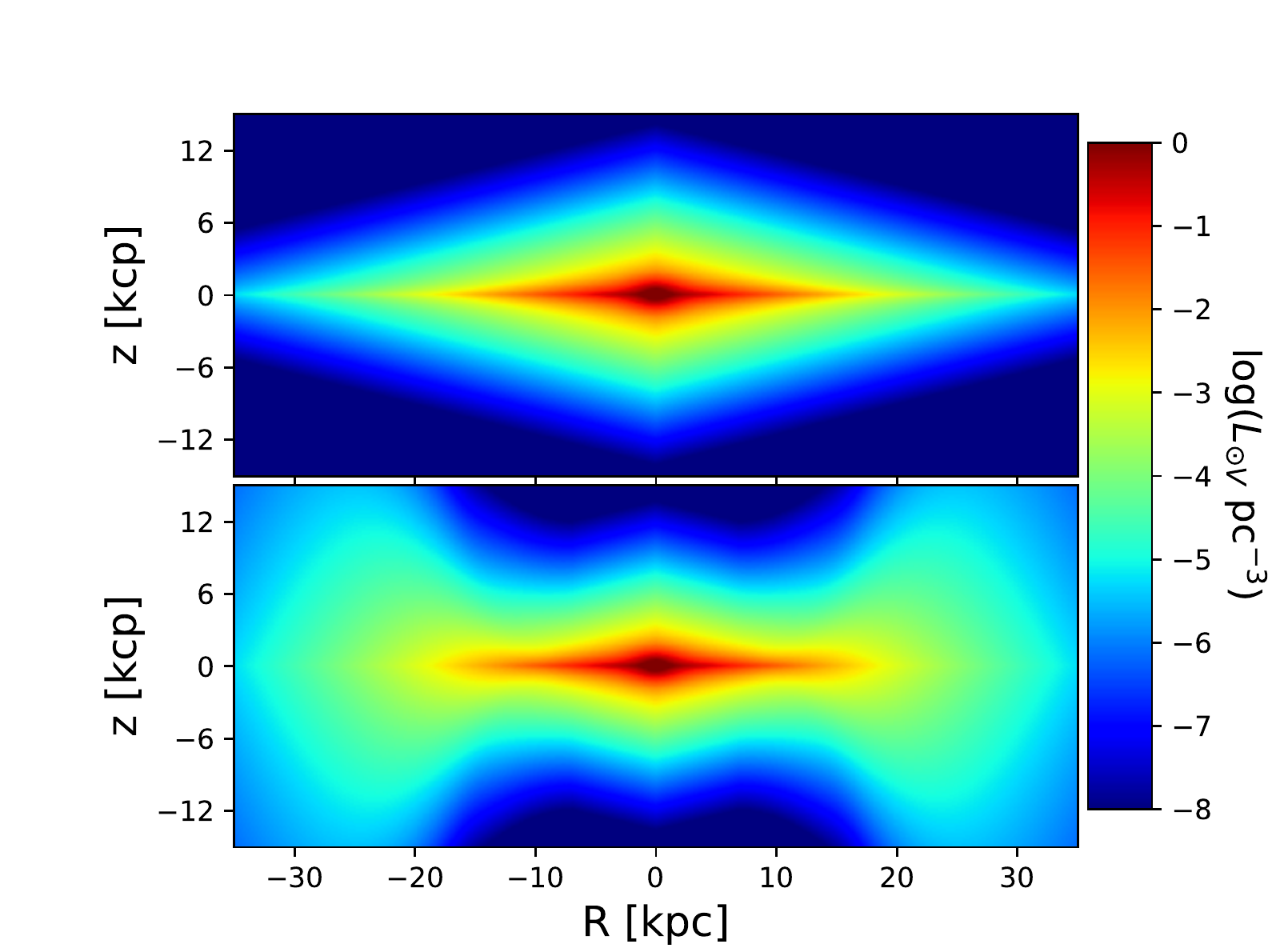}
 \caption{\textbf{Luminosity density profile of the Milky Way without and with flaring.} The top panel shows the stellar luminosity density for our main model described in Section~\ref{subsec: MW main}. The bottom panel shows the flared Milky Way model (see Section~\ref{subsec: MW flared}). The x-axis is the radial direction, $R$, reaching up to $\pm$ 35~kpc away from the Galactic Center at (0, 0)~kpc and the y-axis is the displacement from the midplane, $z$, ranging up to $\pm$ 15~kpc. The colormap is the logarithmic luminosity density.}\label{im:mW dens}
\end{figure}

Since the flare starts only at larger radii, we assume the scale height also for the thin disk to be constant up to the solar position. For $R < R_{\odot}$ $ H_{\rm thin} = H_{\rm thin}(R_{\odot})$ and for $R < R_{\rm ft}$ $ H_{\rm thick} = H_{\rm thick}(R_{\rm ft})$. We use as values for the scale heights at solar position, the values from \citet{juric08}, $H_{\rm thin}(R_{\odot})$ = 300~pc and $H_{\rm thick}(R_{\odot})$ = 900~pc. See Figure~\ref{im:mW dens} bottom panel for the density profile with flare.

\subsection{MGE with negative Gaussians but non-negative density everywhere}

\begin{figure}
    \centering
    \includegraphics[width=\columnwidth]{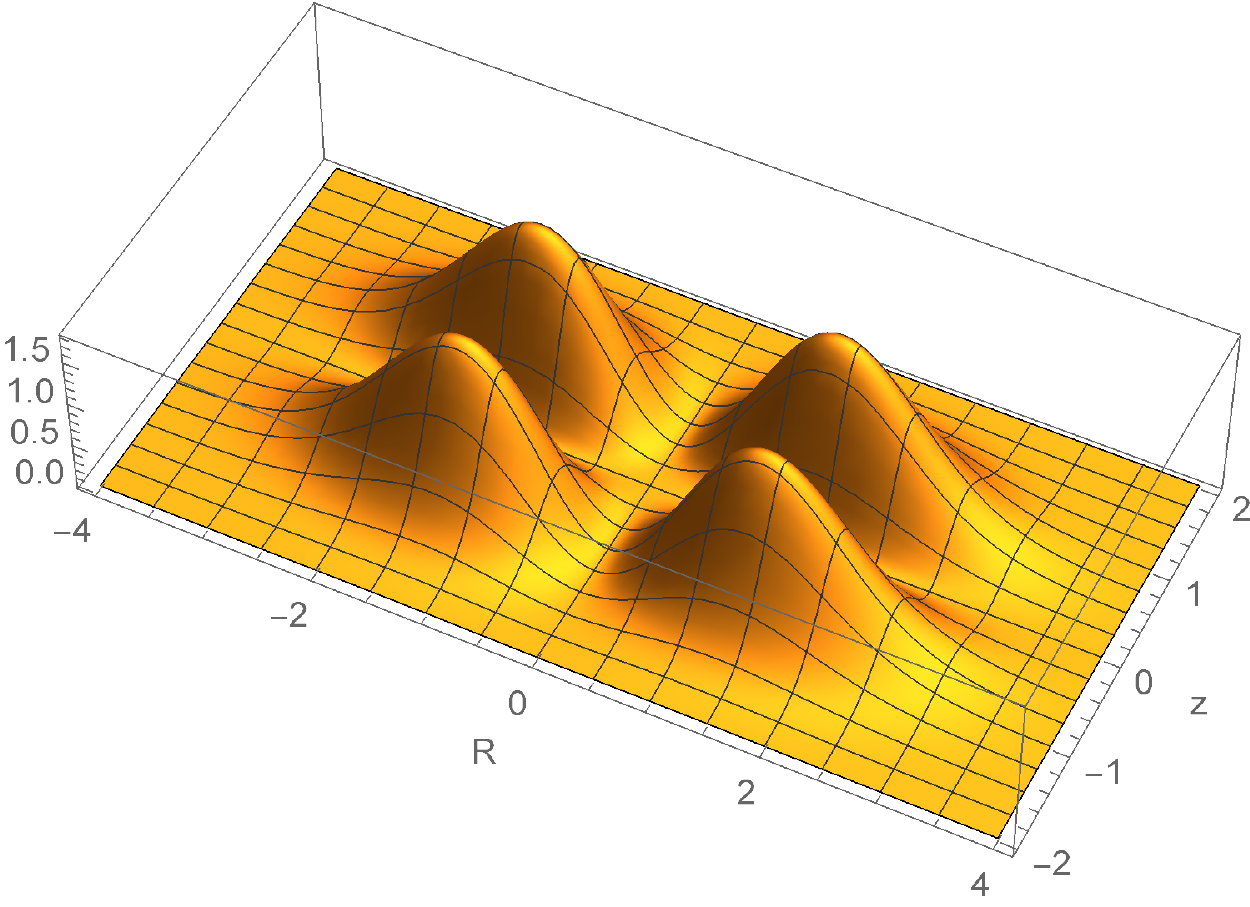}
    \caption{{\bf New MGE basis function.} This basis function consists of four positive/negative Gaussians centred on the origin. When combined with the usual positive Gaussians, allows for extra flexibility in the MGE fitting, while still guaranteeing a non-negative density everywhere.}
    \label{im:mge_basis}
\end{figure}

An MGE with only positive Gaussians cannot describe a density distribution like the one in Figure~\ref{im:mW dens}. A good fit can be obtained, for a restricted spatial extent, allowing for negative Gaussians (keyword \texttt{negative=True} in \textsc{mge\_fit\_sectors} of the \textsc{MgeFit} package of \citealt{Mge2002}). However, we verified that in this case the resulting MGE has strongly negative density at larger radii and produces clearly nonphysical results when used with \textsc{JAM}. The inability of constraining the density to be positive, and the need to test for the positivity of the total density, is a general problem when allowing for negative Gaussian in MGE fits.

To obtain an acceptable MGE fit to the flared disk, while guaranteeing a positive total density, we needed to make some modifications to the \textsc{mge\_fit\_sectors} procedure. For this, in addition to the usual positive Gaussians basis functions in the MGE, we introduced a set of new basis functions, which we produced by co-adding four positive and negative Gaussians with the same absolute peak value, in such a way that their sum is everywhere positive as follows:
\begin{eqnarray}
    G(R, z) &=& e^{-\frac{1}{2 \sigma^2}\left(\frac{R^2}{f^2}+\frac{z^2}{f^2 q^2}\right)}
    +e^{-\frac{1}{2 \sigma ^2}\left(R^2+\frac{z^2}{q^2}\right)}\nonumber\\
    &&-e^{-\frac{1}{2 \sigma ^2}\left(\frac{R^2}{f^2}+\frac{z^2}{q^2}\right)}
    -e^{-\frac{1}{2 \sigma ^2}\left(R^2+\frac{z^2}{f^2q^2}\right)}.
\end{eqnarray}
Here, $\sigma$ and $q$ are two parameters that change for different MGE components (like the corresponding parameters for the MGE Gaussians), while $f=0.9$ is a constant $f\lesssim1$, which is the same for all basis functions. With this definition, the new MGE basis function is $G(R, z)=0$ along the $R$ or $z$ axes, and presents four bi-symmetric maxima, as illustrated in  Figure~\ref{im:mge_basis}.

The new basis function $G(R,z)$ has an asymptotic behaviour at large radii $G\sim\exp(-r^2)$ for $r\rightarrow\infty$, as the standard MGE,  while  along the axes $G\sim R^2$ for $R\rightarrow0$ and $G\sim z^2$ for $z\rightarrow0$. This implies that, while it can describe decreasing profiles towards the centre, it cannot approximate drops of the central density profile steeper than $\rho\propto r^2$. For this reason, it can only approximate the flared density of Figure~\ref{im:mW dens} down to a certain isophote.

Apart from the introduction of the new basis function, the fit is obtained using the same algorithm described in \citet{Mge2002}, while still enforcing positivity constraints for all basis functions. Once the fit converges, all the 4-Gaussian basis functions are split into their four individual Gaussians components in output, producing an MGE which is composed of both positive and negative Gaussians. In this way, no further changes are required to use the produced MGE model with other software like JAM.

The new MGE has 49 Gaussians, but these come from 28 basis-functions, some of which are constructed from groups of 4-Gaussians, which are fitted as single basis-functions. This new implementation of the MGE fit is/will be included in the new release of the MGE package (footnote~\ref{foot:mge}).

\begin{figure*}
\centering
\includegraphics[width=2.1\columnwidth]{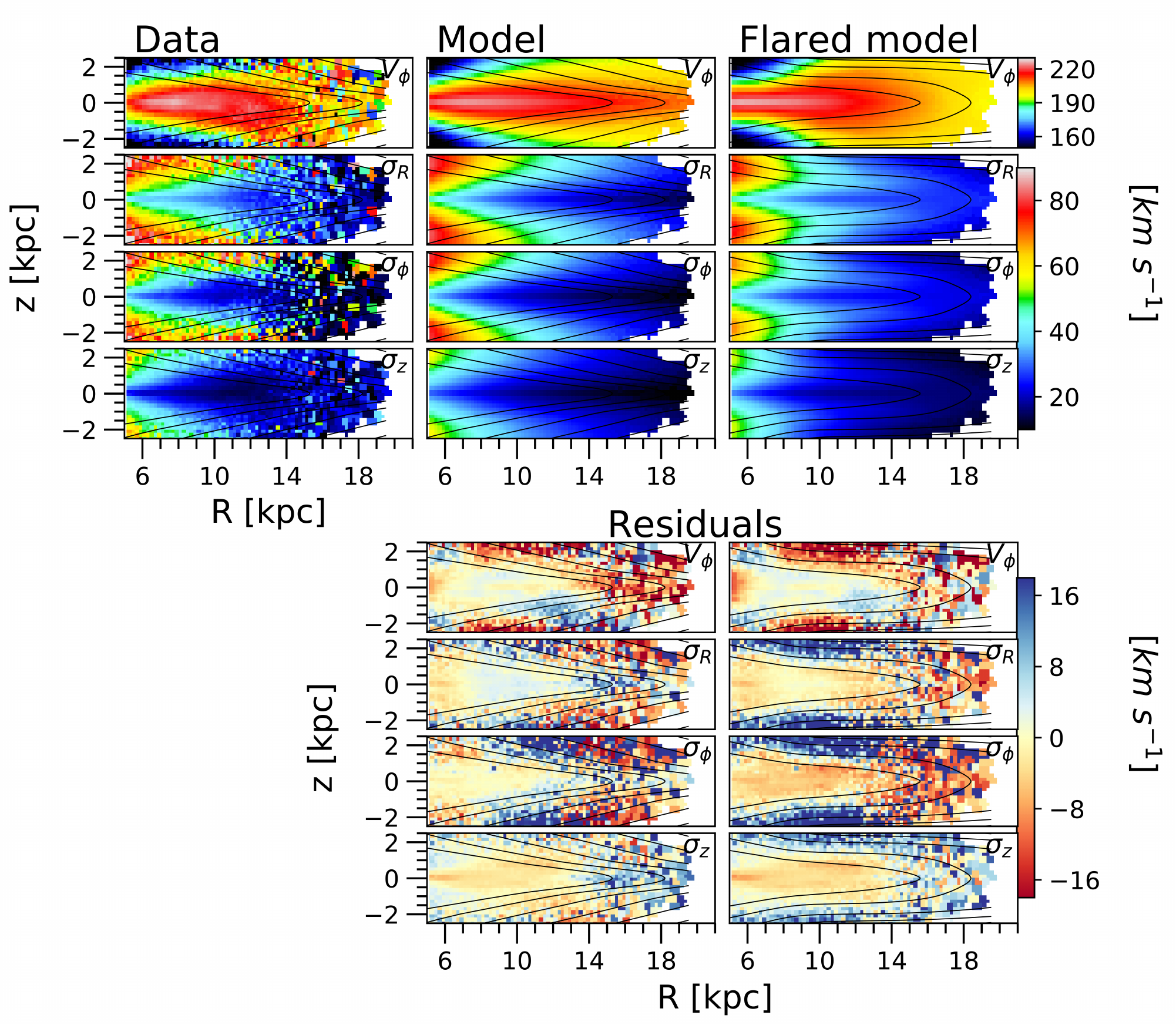}
 \caption{\textbf{Best-fit for the JAM$_{\rm sph}$.} From left to right the columns show the data, the JAM model with the \citet{juric08} stellar density distribution and the JAM model with flaring. At the bottom are the residuals (data-model) for the two JAM models. From top to bottom the rows show $V_{\phi}, \sigma_{R}, \sigma_{\phi}, \sigma_{z}$. The black lines are the stellar density (as described in Section~\ref{subsec: MW main}) contours, except for the flared model which has the contours of the flared disk model described in Section~\ref{subsec: MW flared}.}\label{im:jam sys}
\end{figure*} 

\subsection{Modelling}

As dynamical model we use JAM$_{\rm sph}$ approach\footnote{We use the Python version 6.1.2 of the \textsc{jampy} package available from \url{https://pypi.org/project/jampy/}} \citep{Cappellari2019}, which is based on the solution of the axisymmetric Jeans equations and assumes a spherically aligned velocity ellipsoid \citep{Bacon83, Bacon85}. The spherical alignment has proven to describe the \textit{Gaia} data in the outer halo \citep{wegg19} and in the disk region \citep{Hagen2019, Everall2019}. This was also confirmed by the results in \citetalias{Nitschai2020}.

In Figure~\ref{im:vel map} we have drawn a red square around the data we will use for our model: we only fit the axisymetric in-plane kinematics ($v_\phi, \sigma_r, \sigma_\phi, \sigma_z$). Neither vertical motion, nor tracer distribution are fit. Due to the steady state assumption of our model, $v_{R}$ and $v_{z}$ are assumed to be zero and therefore they do not need to be fitted. The deviations from zero in the data due to non-axisymmetric signatures \citep{Antoja2018, Gaia2018_nonaxi} are small enough for the purpose of our model.

The features which we do not include in our stellar distribution of the Galaxy are the bar in the inner region and warp in the outer disk since both signatures are non-axisymmetric and we cannot model them with JAM. In particular, the bar dominates within 5 kpc from the Galactic Centre \citep{Wegg15, Bovy19}, but as previously mentioned (see Section~\ref{subsec: vel maps}, cut at 5~kpc) we remove all the stars in the central area for our model. For the outer disk one would need to add a measurement of the warp, $z_w(R, \phi)$, \citep[e.g.][]{Li2019} to the stellar distribution. However, since our model is axisymmetric, we cannot add this to our model. In future works one could investigate if a shift in $z$ direction would be a possible parametrization of the warp for an axisymmetric model like ours.

For our modelling we keep the scale radius of the dark matter halo, $r_s$, constant at 14.8~kpc \citep{eilers18}, since we know that it is not constrained well with our JAM model (\citetalias{Nitschai2020}). In addition, 14.8~kpc agrees with literature values ranging from ($\sim$ 10 to 20~kpc). Further, because our data are limited to the disk region, they are not strongly constraining the dark matter axial ratio. Hence, we keep $q_{\rm DM}$ fixed to 1.3 \citep{Posti19}, which they derive using globular clusters. See Appendix~\ref{ap: q} for a model with free $q_{\rm DM}$.

The standard model has 7 free parameters: (i) the inner logarithmic slope of the dark matter halo ($\alpha_{\rm DM}$); (ii) the dark matter fraction within a sphere with radius R$_{\odot}$ ($f_{\rm DM}$); (iii-vi) the velocity dispersion ratios or anisotropies ($\sigma_{\theta}/\sigma_{r}$ and  $\sigma_{\phi}/\sigma_{r}$) for both the flattest ($q_{\rm MGE}<0.2$) Gaussian components (subscript 1) of the MGE and the rest (subscript 2); and (vii) the mass-to-light ratio of the stellar component in the $V$-band [$(M_{\ast}/L)_{V}$].

The fit to the data is performed using the \textsc{emcee} python package of \citet{emcee}, which implements the affine invariant ensemble sampler for Markov chain Monte Carlo (MCMC) proposed by \citet{Goodman2010}.

\section{Results} \label{sec:res}

\subsection{Exponential disk model}\label{subsec: main}

Using the data and uncertainties described in Section~\ref{sec:data} we perform MCMC fits, for the Milky Way distribution without flaring as described in Section~\ref{subsec: MW main}.

The best fitting model, gives a $\chi^2 \sim 9.2$ which is much larger than 1 that we want. This is mainly because there are systematic uncertainties due to non-axisymmetric features, spiral structure, the equilibrium assumption of JAM and the choice of the tracer distribution, which our model cannot take into account. This explains also why we get very small statistical errors for our free parameters. The systematic errors would dominate here and we can use this result to get an estimate of the systematic uncertainties, which are otherwise impossible to determine.

\begin{deluxetable*}{cCCC}
\tablecaption{Best-fitting parameters \label{tab:resuts}}
\tablewidth{0pt}
\tablehead{
\colhead{Parameters} & \colhead{Model} & \colhead{Free anisotropies} & \colhead{Flared Model}}
\startdata
 $\alpha_{\rm DM}$  & -1.602 \pm 0.015_{\rm stat} \pm 0.079_{\rm syst} & -1.592\pm 0.029_{\rm stat} \pm 0.079_{\rm syst} & -1.574 \pm 0.023_{\rm stat} \\
 $f_{\rm DM}$ & 0.811\pm 0.006_{\rm stat} \pm 0.014_{\rm syst} & 0.819\pm 0.013_{\rm stat} \pm 0.014_{\rm syst} &  0.663\pm 0.007_{\rm stat} \\
 $(\sigma_{\theta}/\sigma_{r})_{1}$ & 0.663\pm 0.006_{\rm stat}\pm 0.021_{\rm syst} & - & 0.610 \pm 0.0003_{\rm stat}\\
 $(\sigma_{\theta}/\sigma_{r})_{2}$ & 0.577\pm 0.011_{\rm stat} \pm 0.038_{\rm syst} & - & 0.632 \pm 0.003_{\rm stat}\\
 $(\sigma_{\phi}/\sigma_{r})_{1}$ & 0.711\pm 0.016_{\rm stat} \pm 0.028_{\rm syst} & - & 0.848 \pm 0.007_{\rm stat}\\
 $(\sigma_{\phi}/\sigma_{r})_{2}$ & 1.001\pm 0.019_{\rm stat} \pm 0.054_{\rm syst} & - & 0.853\pm  0.006_{\rm stat}\\
 $(M_{\ast}/L)_{V}$ & 0.413\pm 0.012_{\rm stat} \pm 0.031_{\rm syst} & 0.396\pm 0.028_{\rm stat} \pm 0.031_{\rm syst} & 0.715\pm 0.015_{\rm stat} \\
 $\chi^{2}_{\rm DOF}$ & 0.91 & 0.81 & 1.3\\
 $v_{\rm circ}(R_{\odot})$ [~km~s$^{-1}$] & 234.7\pm 0.3_{\rm stat} \pm 1.7_{\rm syst} & 234.6\pm 0.4_{\rm stat} \pm 1.7_{\rm syst} & 234.6\pm 0.24_{\rm stat} \\
 $a_{\rm circ}$[~km~s$^{-1}$~kpc$^{-1}$] & -1.78\pm 0.05_{\rm stat} \pm 0.34_{\rm syst} & -1.69\pm 0.08_{\rm stat} \pm 0.34_{\rm syst} & -2.39 \pm 0.05_{\rm stat}\\
\enddata
\tablecomments{The uncertainties given in this table as statistical errors are derived from the posterior distributions and are the formal errors. For the estimate of the systematic uncertainties we also used the values from Table~\ref{tab:4 parts} in Section~\ref{subsec: 4 sectors} and we estimate them to be half of the difference between minimum and maximum of all our models with an exponential disk without flaring.}
\end{deluxetable*}

To get more reasonable errors, we check the precision to which our model explains the data. To do this we add quadratically a value in km~s$^{-1}$ to our bootstrapped errors:
\begin{equation}
    \mathrm{errors}_i^2 = \mathrm{errors}_{\mathrm{bootstr.,} i}^2 + \mathrm{errors}_{\mathrm{s.,} i}^2.
\end{equation}
Where $\mathrm{errors}_{\mathrm{s.,} i}$ such that we get a $\chi^2 = 1$ for each component $i = (v_{\phi}, \sigma_{R}, \sigma_{\phi}, \sigma_{z}$). The added values for this model are (6.03, 5.01, 6.55, 3.32)~km~s$^{-1}$ respectively. With these higher errors for the data we do a new MCMC fit and the best-fitting values are listed in the second column (`Model') of Table~\ref{tab:resuts}. 

However, the statistical uncertainties from the posterior of our the best-fitting parameters are only formal errors, which are quite small also because of the large number of stars that we use and systematic uncertainties dominate. To estimate our systematics we investigate the axisymmetry assumption of our model in Section~\ref{subsec: 4 sectors} and we give the systematic errors as half of the difference between minimum and maximum of all our models with an exponential disk without flaring.

The posterior distribution is shown in the Appendix~\ref{ap: posterior} Figure~\ref{im: posterior sys} and the JAM model is shown in the second column of Figure~\ref{im:jam sys}.

\begin{figure*}
\centering
\includegraphics[width=1.7\columnwidth]{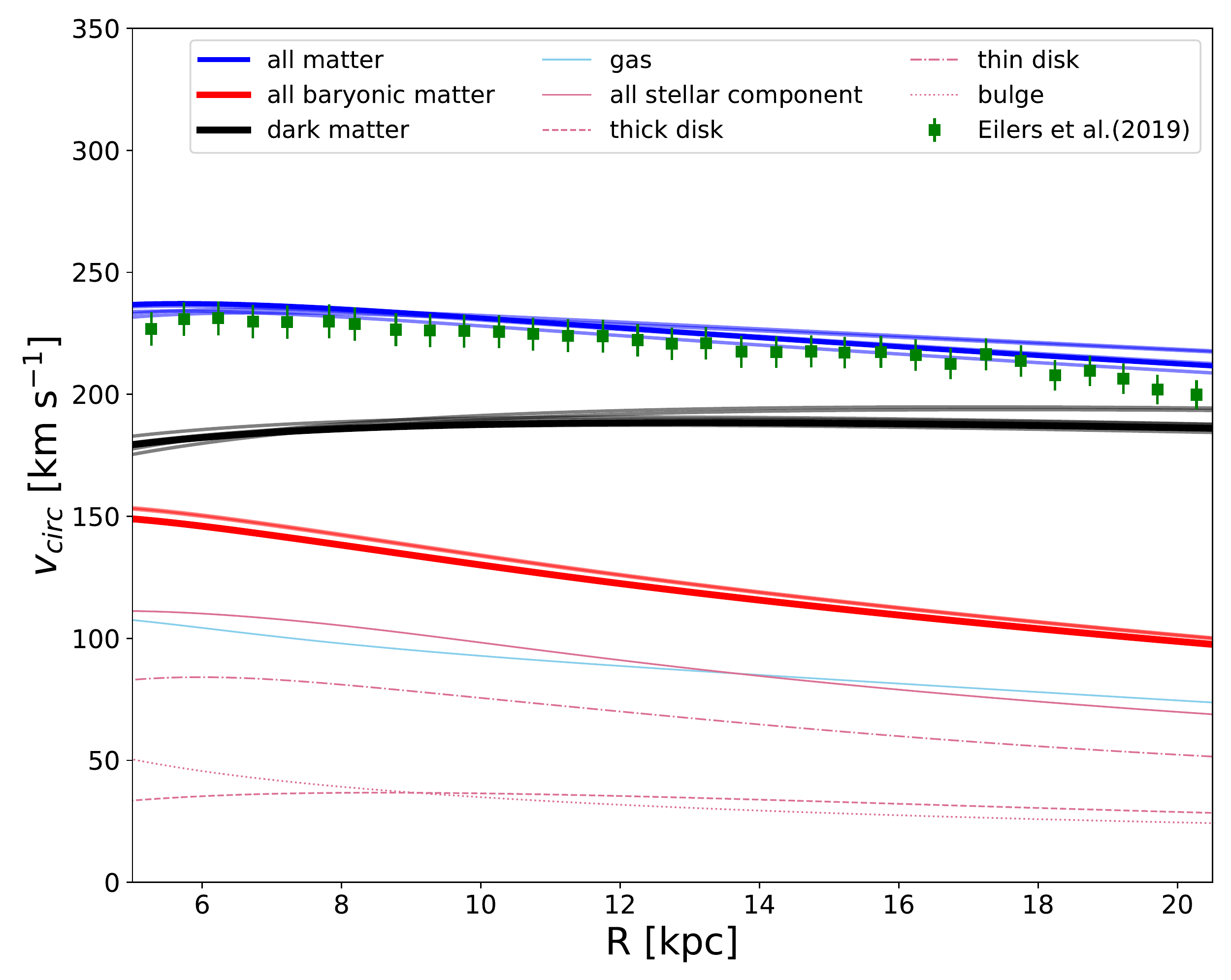}
 \caption{\textbf{Circular velocity for the JAM$_{\rm sph}$ model without flaring.} The blue solid line is our total circular velocity from our best fit, the green squares are the measurements by \citet{eilers18} with systematic uncertainties of 3 per cent, the dark matter contribution is shown as a black solid line and the baryonic contribution as a red solid line. For each of this three lines there are five fainter lines (close around them), which are the best-fitting results from the four different sectors of the data (Section~\ref{subsec: 4 sectors}) and the model with free anisotropies and they give as an estimate of the systematic uncertainty. Only for the best-fitting solution of our main model without flaring we also have plotted the gas contribution as a light blue line and the total stellar contribution as a light violate solid line, while this can be separated further into the bulge (light violate dotted line), the thin (light violate dashed-dotted line) and thick disk (light violate dashed line). The contribution of the different components plotted here,depends on the assumed stellar and dark matter density profiles, and with our model we only constrain tightly the total circular velocity.}
\label{im:vcirc sys}
\end{figure*}

The circular velocity is plotted in Figure~\ref{im:vcirc sys} and has a value of $(234.7 \pm 0.3_{\rm stat} \pm 1.7_{\rm syst})$~km~s$^{-1}$ at the solar radius. The slope of the circular velocity curve between 6.2~kpc and 20.2~kpc, which is the region were it can be approximated by a straight line, is declining at $(-1.78\pm 0.05_{\rm stat} \pm 0.34_{\rm syst})$~km~s$^{-1}$~kpc$^{-1}$. This is consistent with the results of \citet{eilers18}. A test to investigate the small offset between our and \citet{eilers18} measurements of the circular velocity can be found in the Appendix~\ref{ap:circ}. In Figure~\ref{im:vcirc sys} we also show the contribution of the different components to the total circular velocity according to our best fit. However, one has to note that because the covariance of the mass-to-light ratio, the dark matter fraction, and the dark matter slope (see Figure~\ref{im: posterior sys}), which is expected, it is not the only possible decomposition. One could decrease the dark matter contribution while increasing the stellar component and still get the same total result, which is what we constrain. The smaller than NFW ($ \alpha_{\rm DM} = -1$) dark matter slope also increases the dark matter contribution towards smaller radii inside $R_{\odot}$. This makes the dark matter contribution more dominant for our circular velocity at small radii in comparison to other works \citep[e.g.][and what we see in Appendix~\ref{ap:circ} Figure~\ref{im:circ eilers}]{eilers18} where usually a NFW profile is assumed. Hence, we only tightly constrain the total circular velocity and not the different components of it, which depends on the assumed stellar and dark matter density profiles.

One should also note that even though we use giant stars, we do not take the asymmetric drift into account, since we use the full Jeans equations with JAM. The circular velocity is a result of the gravitational potential that we get from our best fit to the individual velocities.

Additionally, for the best fitting model the dark matter density at the solar radius is $\rho_{\rm DM}(R_{\odot}) = (0.00892 \pm 0.00007_{\rm stat} \pm 0.00056_{\rm syst})$~M$_{\odot}$~pc$^{-3}$ and the total density $\rho_{\rm tot}(R_{\odot}) = (0.0672\pm 0.0006_{\rm stat} \pm 0.0015_{\rm syst})$~M$_{\odot}$~pc$^{-3}$. The total density logarithmic slope in the range of $5\leq R\leq19.5$~kpc is $\alpha_{\rm tot} = -2.367 \pm 0.007_{\rm stat} \pm 0.047_{\rm syst}$. All statistical errors are the formal errors from the posterior distribution and do not include the systematic uncertainties that would increase the errors.

Comparing these results to our previous work \citepalias{Nitschai2020}, the JAM model agrees with our previous results and all parameters agree within the respective errors. Specifically, the total density, and in particular on the total slope and dark matter fraction, have smaller uncertainties and are more reliable, because we have a longer radial baseline, and more data. The discrepancy between the total logarithmic density slope of our best fitting model and that from \citetalias{Nitschai2020} can be explained due to the different radial range probed ($3.5\leq R\leq 12.5$~kpc) and the different $r_s$ (20~kpc) assumed for the dark matter halo. If we constrain our slope to a similar range as in \citetalias{Nitschai2020}, our value would decrease to $-2.278 \pm 0.007_{\rm stat} \pm 0.047_{\rm syst}$ and would get even smaller for a scale radius of 20~kpc. So in a similar range our finding for the total density slope is also consisted with that of \citetalias{Nitschai2020}.

\begin{figure*}
\centering
\includegraphics[width=1.7\columnwidth]{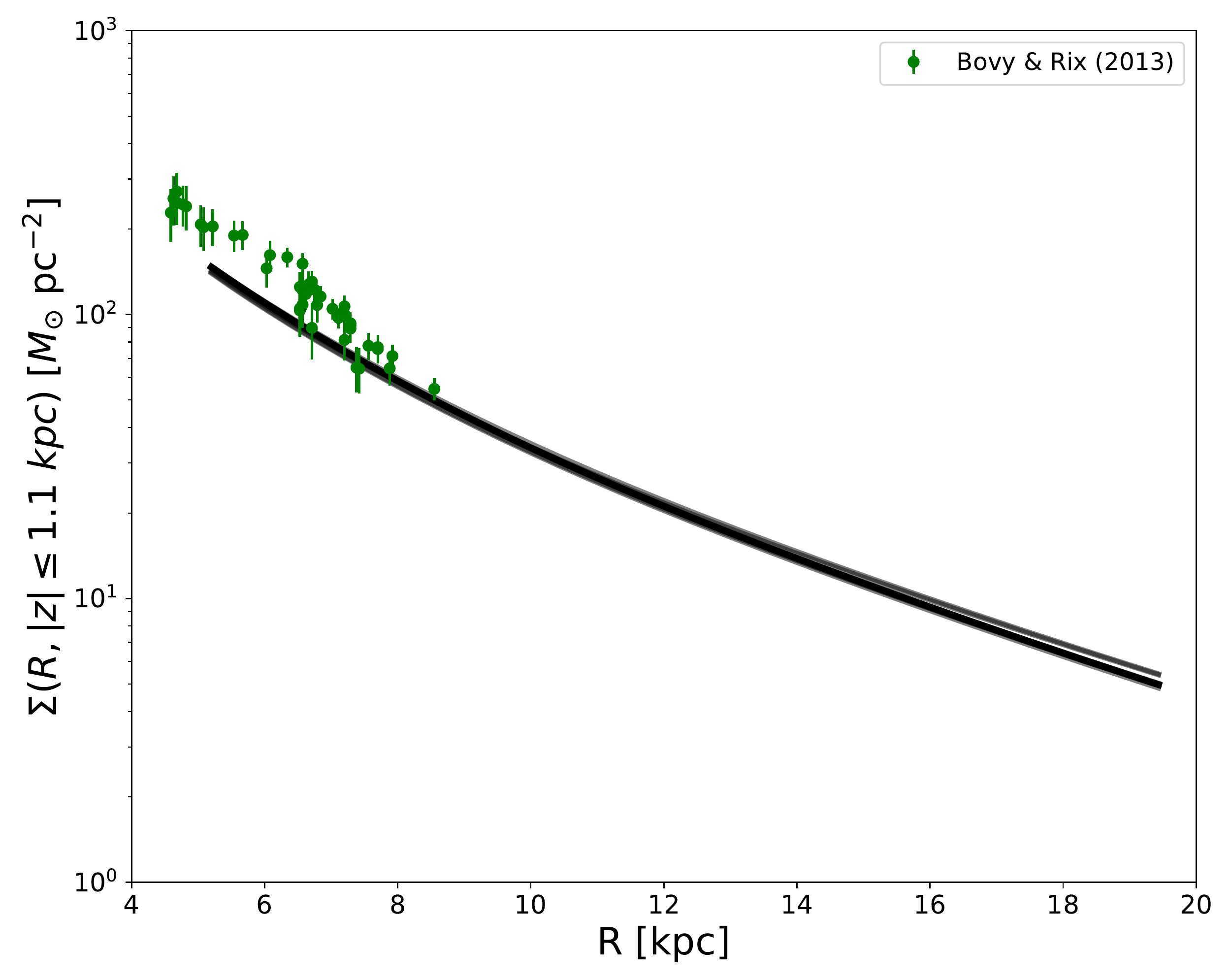}
 \caption{\textbf{Surface density for the JAM$_{\rm sph}$ model without flaring.} The total surface density is calculated between -1.1~kpc $\leq z \leq$ 1.1~kpc for our best-fitting model. The green dots are the values from \citet{Bovy13} and the five fainter black lines (close around it) are calculated with the best-fitting results from the four different sectors of the data (Section~\ref{subsec: 4 sectors}) and the model with free anisotropie. The radial range (5~kpc to 19.1~kpc) of the surface density in this plot, is between the minimum and maximum radius of the data which we used for our best-fitting model (see Figure~\ref{im:jam sys}).}
\label{im:surface dens}
\end{figure*}

In addition, we also have calculated the total surface density as a function of radius by analytically integrating the MGE Gaussians. The integral of one Gaussian, between the cylindrical radii $R_1$ and $R_2$ and within the height range $|z|<z_{\rm max}$ is given by the following expression
\begin{equation}
M = (2\pi)^{3/2} \rho_0 q \sigma ^3 \left(e^{-\frac{R_1^2}{2
   \sigma ^2}}-e^{-\frac{R_2^2}{2 \sigma ^2}}\right)
   \text{erf}\left(\frac{z_{\rm max}}{q \sigma\sqrt{2}}\right).
\end{equation}
This integral was performed for all the MGE Gaussians for the stellar, gas and dark matter components for -1.1~kpc $\leq z \leq$ 1.1~kpc for direct comparison with \citet{Bovy13}. The resulting surface density is shown in Figure~\ref{im:surface dens}. Our surface density is a little bit lower for radii smaller than $R_{\odot}$ but agrees within the uncertainties at the solar radius, $\Sigma(R{_\odot}, |z|\leq$ 1.1~kpc) = $(55.5 \pm 0.5_{\rm stat} \pm 1.7_{\rm syst})$~M$_{\odot}$~pc$^{-2}$. Moreover, \citet{Piffl2014} give for the solar radius $\Sigma(|z|\leq$ 0.9~kpc) = (69 $\pm$ 15)~M$_{\odot}$~pc$^{-2}$ and our value of $\Sigma(R{_\odot}, |z|\leq$ 0.9~kpc) = $(51.1 \pm 0.4_{\rm stat} \pm 1.7_{\rm syst})$~M$_{\odot}$~pc$^{-2}$ lies within their 3$\sigma$ range. Since also our total surface density agrees within the uncertainties with previous findings, it gives us another confirmation that our model constrains well the total density, circular velocity, and the potential, and only the exact decomposition into the different components (dark matter, stellar and gas), which is not the scope of this work, is not tightly constrained.

Further, we also performed a fit allowing the 18 Gaussians in the MGE model of the stellar component, to have a different anisotropy value, $(\sigma_{\theta}/\sigma_{r})$ and  $(\sigma_{\phi}/\sigma_{r})$. This model has a total of 39 free parameters and the ones that are not the anisotropies are listed in column `Free anisotropies' in Table~\ref{tab:resuts}. This model gives a better fit to the data, since it has more parameters but all `interesting'  parameters agree within the 3$\sigma$ error range with our main model.

\subsection{Assessing systematic errors due to non-axisymmetries}\label{subsec: 4 sectors}

Our Galaxy contains non-axisymmetric features due to the Galactic bar \citep[e.g.][]{monari16, khoperskov19}, spiral arms \citep[e.g.][]{reid19, eilers20}, the warp of the disk \citep[e.g.][]{vazquez08, li20}, or interactions and mergers with satellite galaxies \citep[e.g.][]{helmi18, koppelman19}. 
We now investigate how the model assumption of axisymmetry affects our results by dividing the data set into four sectors, depending on positive or negative $z$ and $\phi$: [$z \geq$ 0~kpc, $\phi \geq$ 0$^\circ$], [$z \geq$ 0~kpc, $\phi<$ 0$^\circ$], [$z<$ 0~kpc, $\phi \geq$ 0$^\circ$] and [$z<$ 0~kpc, $\phi<$ 0$^\circ$]. If the Milky Way was perfectly axisymmetric, our axisymmetric model should give the same results within the errors for each of the four sectors. Hence, we now fit for each sector separately and compare the different results, in order to estimate how much the non-axisymmetric features of our Galaxy influence our model and estimate the systematic uncertainties of the free parameters.

\begin{deluxetable*}{cCCCCCC}
\tablecaption{Best-fitting parameters for the four sectors \label{tab:4 parts}}
\tablewidth{0pt}
\tablehead{
\colhead{Parameters} & \colhead{$z\geq$0~kpc, $\phi\geq$ 0$^\circ$} & \colhead{$z\geq$0~kpc, $\phi<$ 0$^\circ$} & \colhead{$z<$0~kpc, $\phi\geq$ 0$^\circ$} & \colhead{$z<$0~kpc, $\phi<$ 0$^\circ$} & \colhead{Mean} & \colhead{Syst. error}}
\startdata
 $\alpha_{\rm DM}$ & -1.623 \pm 0.017_{\rm stat} & -1.627\pm 0.018_{\rm stat} & -1.469\pm 0.019_{\rm stat} & -1.487 \pm 0.019_{\rm stat} & -1.552\pm 0.079 & \pm 0.079\\
 $f_{\rm DM}$ & 0.837\pm 0.007_{\rm stat} & 0.830 \pm 0.008_{\rm stat} & 0.816\pm 0.007_{\rm stat} & 0.838\pm 0.007_{\rm stat} & 0.830\pm 0.011 & \pm 0.014\\
 $(\sigma_{\theta}/\sigma_{r})_{1}$ & 0.653\pm 0.006_{\rm stat} & 0.655\pm 0.008_{\rm stat} &  0.666\pm 0.008_{\rm stat} & 0.709\pm 0.010_{\rm stat} & 0.671\pm 0.021 & \pm 0.021\\
 $(\sigma_{\theta}/\sigma_{r})_{2}$ & 0.569\pm 0.011_{\rm stat} & 0.610\pm 0.014_{\rm stat} & 0.601\pm 0.013_{\rm stat} & 0.568\pm 0.014_{\rm stat} & 0.587\pm 0.038 & \pm 0.038\\
 $(\sigma_{\phi}/\sigma_{r})_{1}$ & 0.695\pm 0.017_{\rm stat} & 0.702\pm 0.017_{\rm stat} & 0.702\pm 0.018_{\rm stat} & 0.771\pm 0.019_{\rm stat} & 0.718\pm 0.028 & \pm 0.028\\
 $(\sigma_{\phi}/\sigma_{r})_{2}$ & 0.911\pm 0.021_{\rm stat} & 0.983\pm 0.024_{\rm stat} & 0.940\pm 0.023_{\rm stat} & 0.894\pm 0.024_{\rm stat} & 0.932\pm 0.045 & \pm 0.054\\
 $(M_{\ast}/L)_{V}$ & 0.355\pm 0.015_{\rm stat} & 0.361\pm 0.016_{\rm stat} & 0.401\pm 0.015_{\rm stat} & 0.352\pm 0.015_{\rm stat} & 0.367\pm 0.025 & \pm 0.031\\
 $\chi^{2}_{\rm DOF}$ & 0.98 & 0.97 & 1.03 & 1.05 & - & -\\
 $v_{\rm circ}(R_{\odot})$ [~km~s$^{-1}$] & 233.7\pm 0.3_{\rm stat} & 231.4\pm 0.3_{\rm stat} & 234.3\pm 0.3_{\rm stat} & 232.9\pm 0.3_{\rm stat} & 233.1\pm 1.5 & \pm 1.7\\
 $a_{vcirc}$[~km~s$^{-1}$~kpc$^{-1}$] & -1.71\pm 0.06_{\rm stat} & -1.76\pm 0.07_{\rm stat} & -1.18\pm 0.07_{\rm stat} & -1.11\pm 0.06_{\rm stat} & -1.44\pm 0.33 & \pm 0.34\\
\enddata
\tablecomments{The uncertainties given in this table as statistical errors are derived from the posterior distributions and are the formal errors. For the estimate of the systematic uncertainties we used these values and from Table~\ref{tab:resuts}, in addition we give the mean value of the four sectors and its error, which is half of the difference of the minimum and maximum of the individual sectors (error = (max - min) / 2).}
\end{deluxetable*}

For these fits we use the same errors as for the fit to all data points. The resulting circular velocities are plotted in Figure~\ref{im:vcirc sys} as fainter and thinner blue lines and all the fit parameters are listed in Table~\ref{tab:4 parts}. The results show us that there are some differences depending on which part of the Galaxy we model, however most of the free parameters are within the 3$\sigma$ range of each other and with the total model. Higher discrepancies exist between $\alpha_{\rm DM}$, which are outside the 3$\sigma$ range of the statistical errors.

We can use the discrepancies between the parameters for the different sectors to get an estimate for the systematic uncertainty of our model because of non-axisymmetric features in the data. The mean values in Table~\ref{tab:4 parts} is the mean of the individual sectors and the errors are half of the differences between the maximum and the minimum of the individual sectors. The last column of the table is the total systematic errors = (max-min)/2, including the best fitting values from Table~\ref{tab:resuts} without flaring and the model with free anisotropies. These systematic errors from the sectors and from all our models without flaring might be the same if the minimum and maximum value are from the four sectors.

Additionally, for the four sectors the mean dark matter density at the solar radius is , $\rho_{\rm DM}(R_{\odot}) = (0.00935 \pm 0.00056_{\rm syst}) $~M$_{\odot}$~pc$^{-3}$ and the mean total density, $\rho_{\rm tot}(R_{\odot}) = (0.0654 \pm 0.0015_{\rm syst}) $~M$_{\odot}$~pc$^{-3}$. The mean total density logarithmic slope, $\alpha_{\rm tot} = -2.322 \pm 0.047_{\rm syst}$ for $5\leq R\leq19.5$~kpc and the mean local surface density $\Sigma(R{_\odot}, |z|\leq$ 1.1~kpc) = $(54.6 \pm 1.7_{\rm syst})$~M$_{\odot}$~pc$^{-2}$. All the errors given here are the same systematic errors as in Section~\ref{subsec: main} using the four sectors, the best-fitting model, and the model with free anisotropies.

\subsection{Flared Disk Model}\label{subsec: flared}

In Section~\ref{subsec: MW flared} we also explained how we can change our stellar distribution to account for the flaring of the disk. Using this distribution we test how that affects our model.

We use the same errors as before and the fit results are listed in Table~\ref{tab:resuts} with their formal errors. The model result is shown in the third column of Figure~\ref{im:jam sys} and the posterior distribution is shown in the Appendix~\ref{ap: posterior} Figure~\ref{im: posterior flared}. From the posterior distribution and the table one can notice that the statistical uncertainty of $(\sigma_\theta/\sigma_r)_1$ is an order of magnitude smaller than for the rest of the velocity dispersion ratios, even though the fit has converged properly. This confirms that the formal errors are not reliable since they are too small and the systematic uncertainties of the model would dominate.

Further, the dark matter density at the solar radius, $\rho_{\rm DM}(R_{\odot}) = (0.00715 \pm 0.00007_{\rm stat})$~M$_{\odot}$~pc$^{-3}$ and the total density, $\rho_{\rm tot}(R_{\odot}) = (0.08561\pm 0.00073_{\rm stat})$~M$_{\odot}$~pc$^{-3}$. The total density logarithmic slope, $\alpha_{\rm tot} =-2.451\pm 0.010_{\rm stat}$ for $5\leq R\leq19.5$~kpc and the local surface density $\Sigma(R{_\odot}, |z|\leq$ 1.1~kpc) = $(64.1 \pm 0.6_{\rm stat})$~M$_{\odot}$~pc$^{-2}$ .

\begin{figure}
\centering
\includegraphics[width=0.9\columnwidth]{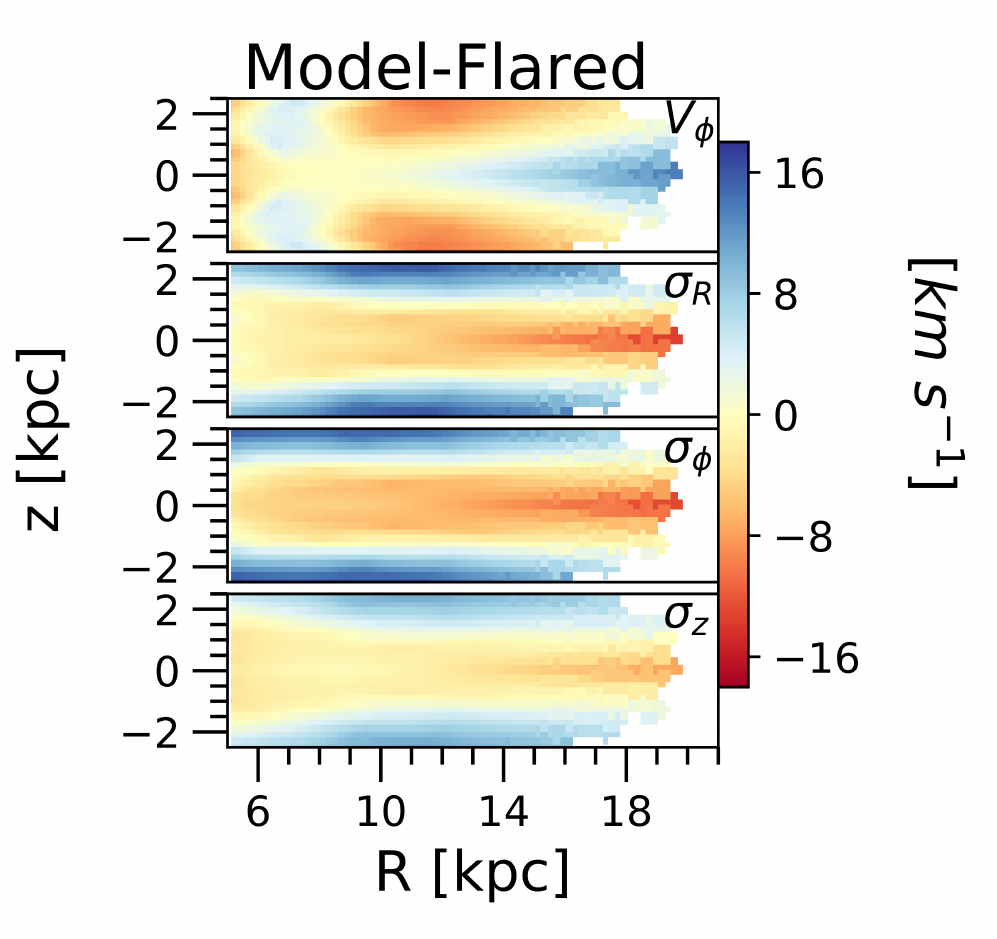}
 \caption{\textbf{The difference between the JAM$_{\rm sph}$ without and with a flared disk.} From top to bottom the rows show $V_{\phi}, \sigma_{R}, \sigma_{\phi}, \sigma_{z}$.}\label{im:flar-model}
\end{figure}

From the $\chi^2$ value one can see that the fit with flaring is not as good as without it, which could be due to our choice of flare parameters. Also, if we compare the residuals of the two models, one can notice that the residuals get higher away from the midplane for the model with flaring. On the other hand, around the midplane it seems to be similarly good or only slightly higher than the model without the flaring. The differences between the models with and without flaring can also be seen in Figure~\ref{im:flar-model}.

We do not have systematic uncertainties because the investigation in the previous section was with a different Milky Way model without flaring and the flared model does not fit the data as well, so we cannot assume that they have the same systematics and further study would be needed. As mentioned above, the flared model might give slightly worse fit to our data only because of the flare parameters we selected, since we just added a parametrization of the scale height, $h(z)$, to our normal model without flaring. Additionally, to fully test if the flared model provides a better fit to the data, one would need to do a detailed study of the flare and test which parametrization for the scale heights is the best for the Milky Way disk. This is beyond the scope of this paper and we just want to get a first idea of how our model changes if we include a flare.

\section{Conclusion} \label{sec:con}

In this paper we construct a dynamical Jeans model of the Milky Way disk using the JAM$_{\rm sph}$ code \citep{Cappellari2019}.

The data we use are based on \textit{Gaia} EDR3, combined with the \citet{Hogg2019} data set, which are combined APOGEE, \textit{Gaia} DR2, 2MASS and WISE data with precise spectrophotometric distance estimates. The combined data set enlarges the range of Galactocentric distances that we cover. We probe an area in range of $5.0 \leq R \leq 19.5$~kpc and $-2.5 \leq z \leq 2.5$~kpc.

Our main results are consistent with \citetalias{Nitschai2020} within the uncertainties, but more reliable because of the more extended radial range of our data.

The best fit value for the dark matter slope is $\alpha_{\rm DM} = -1.602 \pm 0.015_{\rm stat} \pm 0.079_{\rm syst}$ which is smaller than $-1$ (NFW) and the dark matter density at solar radius is $(0.00892 \pm 0.00007_{\rm stat} \pm 0.00056_{\rm syst})$~M$_{\odot}$~pc$^{-3}$. These values agree with our previous work \citepalias{Nitschai2020} and are consistent with other works for the dark matter slope which indicate that a steeper slope than NFW is needed for the disk region \citep{Portail17, Cole17}. The dark matter density at solar position is slightly smaller than previous works \citep[][\citetalias{Nitschai2020}]{McKee2015, mcmillan17}, which might be because we constrain the total density and the decomposition is more uncertain, similar for what we explained for the circular velocity. However, the values still agree in the 3$\sigma$ range of the uncertainties. The circular velocity at the solar position is $(234.7 \pm 0.3_{\rm stat} \pm 1.7_{\rm syst})$~km~s$^{-1}$ with a mild decline towards larger radii of $a_{\rm circ} = (-1.78 \pm 0.05_{\rm stat} \pm 0.34_{\rm syst})$~km~s$^{-1}$~kpc$^{-1}$, which is consistent with the results from \citet{eilers18}. The total density at the solar radius is $\rho_{\rm tot}(R_{\odot}) = (0.0672 \pm 0.0006_{\rm stat} \pm 0.0015_{\rm syst})$~M$_{\odot}$~pc$^{-3}$ and total density logarithmic slope is $\alpha_{\rm tot} = -2.367 \pm 0.007_{\rm stat} \pm 0.047_{\rm syst}$ for $5\leq R\leq19.5$~kpc and $-2.278 \pm 0.007_{\rm stat} \pm 0.047_{\rm syst}$ for $3.5\leq R\leq 12.5$~kpc (as in \citetalias{Nitschai2020}). All these values are consistent our previous work \citepalias{Nitschai2020} within the errors. The total density slope is also consistent with the slope inferred from early-type disk galaxies \citep{Cappellari15} and the total solar density is again smaller compared to other works but agrees within the 3$\sigma$ uncertainties of the work by \citet{McKee2015}. Further, the local surface density is $\Sigma(R{_\odot}, |z|\leq$ 1.1~kpc) = $(55.5 \pm 0.5_{\rm stat} \pm 1.7_{\rm syst})$~M$_{\odot}$~pc$^{-2}$ which is slightly lower than other findings \citep[e.g.][]{Bovy13, Piffl2014} but still within the uncertainties.

Additionally, we also test how non-axisymmetries of the gravitational potential change our result.

We further investigate how a flared disk would change our results. It provides a similarly good fit but in the future one could perform a more thorough analysis of the parameter space to better constrain the flaring of the disk given its dynamical properties.

The regions which are not included in our model are the bar in the inner part and the warp in the outer disk. This features are non-axisymmetric and we cannot reproduce them with our model. We avoid the bar influence by excluding the region inside 5~kpc, but the warp is still effecting our model. Hence, one should note that the regions of our data strongly influenced by this feature (the outer disk) are not well reproduced with our model due to the axisymmetry assumption.

The kinematics used in this paper and the MGE components can
be found as Supplementary data in the online version.
The kinematic data can be used to recreated Figures~\ref{im:vel map}, \ref{im:error} and the best-fitting results of Table~\ref{tab:resuts} while the Kinematics of the individual sectors were used to get the best-fitting parameters of Table~\ref{tab:4 parts}. The MGE datasets are useful for Figures~\ref{im:vel map}, \ref{im:error}, \ref{im:jam sys} and are described in Section~\ref{sec:Method}. The full details are available in the package.

\section*{Acknowledgements}

We thank David W.\ Hogg for his helpful advice and input regarding this project. We thank the anonymous referee for a constructive report which helped to improve the paper.

This work has made use of data from the European Space Agency (ESA) mission {\it Gaia} (\url{https://www.cosmos.esa.int/gaia}), processed by the {\it Gaia}
Data Processing and Analysis Consortium (DPAC,
\url{https://www.cosmos.esa.int/web/gaia/dpac/consortium}). Funding for the DPAC has been provided by national institutions, in particular the institutions
participating in the {\it Gaia} Multilateral Agreement.

N.N. gratefully acknowledges support by the Deutsche Forschungsgemeinschaft (DFG, German Research Foundation) -- Project-ID 138713538 -- SFB 881 (`The Milky Way System', subproject B8).

ACE acknowledges support by NASA through the NASA Hubble Fellowship grant $\#$HF2-51434 awarded by the Space Telescope Science Institute, which is operated by the Association of Universities for Research in Astronomy, Inc., for NASA, under contract NAS5-26555.

\newpage
\appendix
\section{Posterior Distribution} \label{ap: posterior}

\begin{figure*}
\includegraphics[width=\columnwidth]{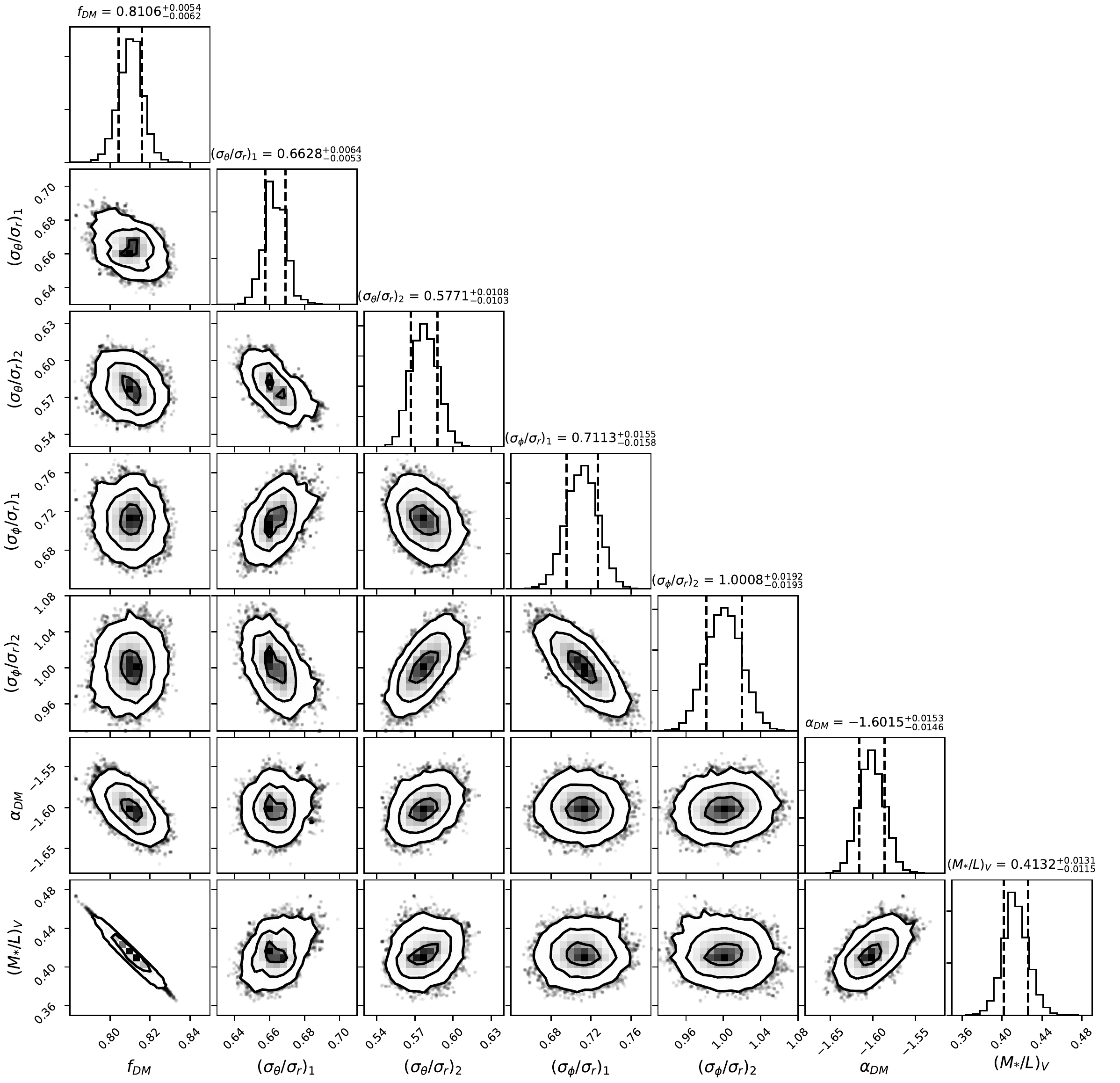}
 \caption{\textbf{Posterior distribution for the best-fitting JAM$_{\rm sph}$ with an exponential disk model.} This is the corner plot for the fit without a flared disk (see Section~\ref{subsec: main}). The panels show posterior probability distributions marginalized over two dimensions (contours) and one dimension (histograms). The thick contours represent
the 1$\sigma$, 2$\sigma$ and 3$\sigma$ confidence levels for one degree of freedom. The numbers with errors on top of each plot are the median and 16th and 84th percentiles of the posterior for each parameter (black dashed lines).}
\label{im: posterior sys}
\end{figure*}

\begin{figure*}
\includegraphics[width=\columnwidth]{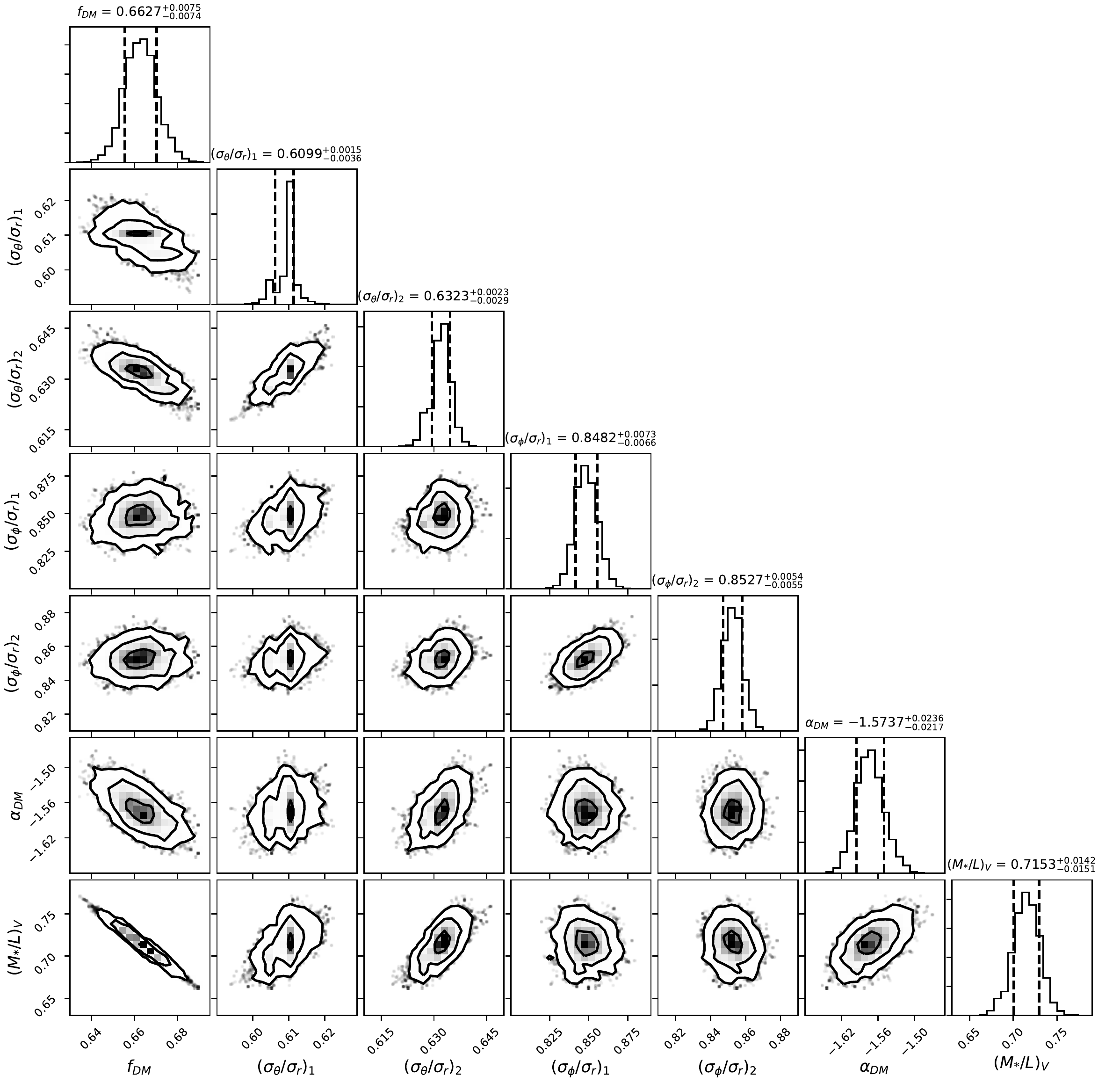}
 \caption{\textbf{Posterior distribution for the best-fitting JAM$_{\rm sph}$ with a flared disk model.} This is the corner plot for the fit with a flared disk (see Section~\ref{subsec: flared}). The panels show posterior probability distributions marginalized over two dimensions (contours) and one dimension (histograms). The thick contours represent
the 1$\sigma$, 2$\sigma$ and 3$\sigma$ confidence levels for one degree of freedom. The numbers with errors on top of each plot are the median and 16th and 84th percentiles of the posterior for each parameter (black dashed lines).}
\label{im: posterior flared}
\end{figure*}

The posterior distribution of the the best-fitting model without flaring (see Section~\ref{subsec: main}) is shown in Figure~\ref{im: posterior sys}. Additionally, Figure~\ref{im: posterior flared} shows the posterior distribution for the best-fitting model with a flared disk (see Section~\ref{subsec: flared}). We have also tested how the result changes with JAM$_{\rm cyl}$ \citep{JAM} for the flared model and we see that the results are almost the same. The only difference is that the small formal errors of $(\sigma_\theta/\sigma_r)_1$ are getting of the same order as the other ratios and that $\chi^2_{\rm DOF}$ is slightly smaller ($\chi^2_{\rm DOF}$ = 1.07), which can be a coincidence since the flared model we use might be wrong, and the cylindrical alignment may happen to compensate for that.

\section{The dark matter axial ratio ($q_{\rm DM}$) as a free parameter}\label{ap: q}

For our main model we have fixed $q_{\rm DM}$=1.3 \citep{Posti19}, here we show a model if we allow $q_{\rm DM}$ to be free. The resulting free parameters with a free $q_{\rm DM}$ for a model without flaring are listed in Table~\ref{tab:q dm}.

What we see also from the posterior distribution, is that $q_{\rm DM}$ tends to get high values and has a correlation with the mass-to-light ratio [$(M_{\ast}/L)_{V}$]. Since, a too high $q_{\rm DM}$ value is nonphysical we keep it fixed for our models in the main text.

\begin{deluxetable}{cC}[h]
\tablecaption{Best-fitting parameters for the model with $q_{\rm DM}$ free \label{tab:q dm}}
\tablewidth{0pt}
\tablehead{
\colhead{Parameters} &  \colhead{$q_{\rm DM}$ free}}
\startdata
 $\alpha_{\rm DM}$ & -1.576\pm 0.026_{\rm stat}\\
 $f_{\rm DM}$ & 0.788\pm 0.009_{\rm stat}\\
 $(\sigma_{\theta}/\sigma_{r})_{1}$ & 0.664\pm 0.007\\
 $(\sigma_{\theta}/\sigma_{r})_{2}$ & 0.572\pm 0.010_{\rm stat}\\
 $(\sigma_{\phi}/\sigma_{r})_{1}$ & 0.707\pm 0.016_{\rm stat}\\
 $(\sigma_{\phi}/\sigma_{r})_{2}$ & 1.004\pm 0.021_{\rm stat}\\
 $(M_{\ast}/L)_{V}$ & 0.480\pm 0.026_{\rm stat}\\
 $\chi^{2}_{\rm DOF}$ & 0.906\\
 $q_{\rm DM}$ & 1.473\pm 0.068_{\rm stat}\\
 $v_{\rm circ}(R_{\odot})$ [~km~s$^{-1}$] & 235.21\pm 0.27_{\rm stat}\\
 $a_{vcirc}$[~km~s$^{-1}$~kpc$^{-1}$] & -1.83\pm 0.05_{\rm stat}\\
\enddata
\tablecomments{All uncertainties given in this table are statistical errors derived from the posterior distributions.}
\end{deluxetable}

\section{Investigating the offset of $V_{\rm circ}$ between our result and previous work}\label{ap:circ}

Additionally, we have investigated the offset that we had found between the circular velocity from \citet{eilers18} and ours. We had already seen an offset in the previous calculation with the best-fitting JAM model in \citetalias{Nitschai2020}.

To do that, we assume the same Milky Way model like in \citet{eilers18}. We adopt a spherical Navarro–Frenk–White profile \citep{NFW96}, for the thin and thick disk we assume Miyamoto–Nagai profiles \citep{Miyamoto75}, and for the bulge we assume a spherical Plummer potential \citep{Plummer_1911}, while adapting the parameter values of \citet[][model I]{pouliasis17}. Furthermore, we use only the \citet{Hogg2019} data that satisfy $|z| <$ 0.5~kpc or $\tan(z/R)<6^\circ$ and and have more than 3 stars in each 200~pc $\times$ 200~pc bin. Finally, we assume R$_{\odot}=8.122$~kpc \citep{GC18}, $z_{\odot}=0.025$~kpc \citep{juric08} and as solar velocities in cylindrical Galactic coordinates $(U_{\odot}, V_{\odot}, W_{\odot})= (-11.1, 245.8, 7.8)$~km~s$^{-1}$ \citep{Reid04}.

\begin{figure}[t]
\centering
 \includegraphics[width=0.7\columnwidth]{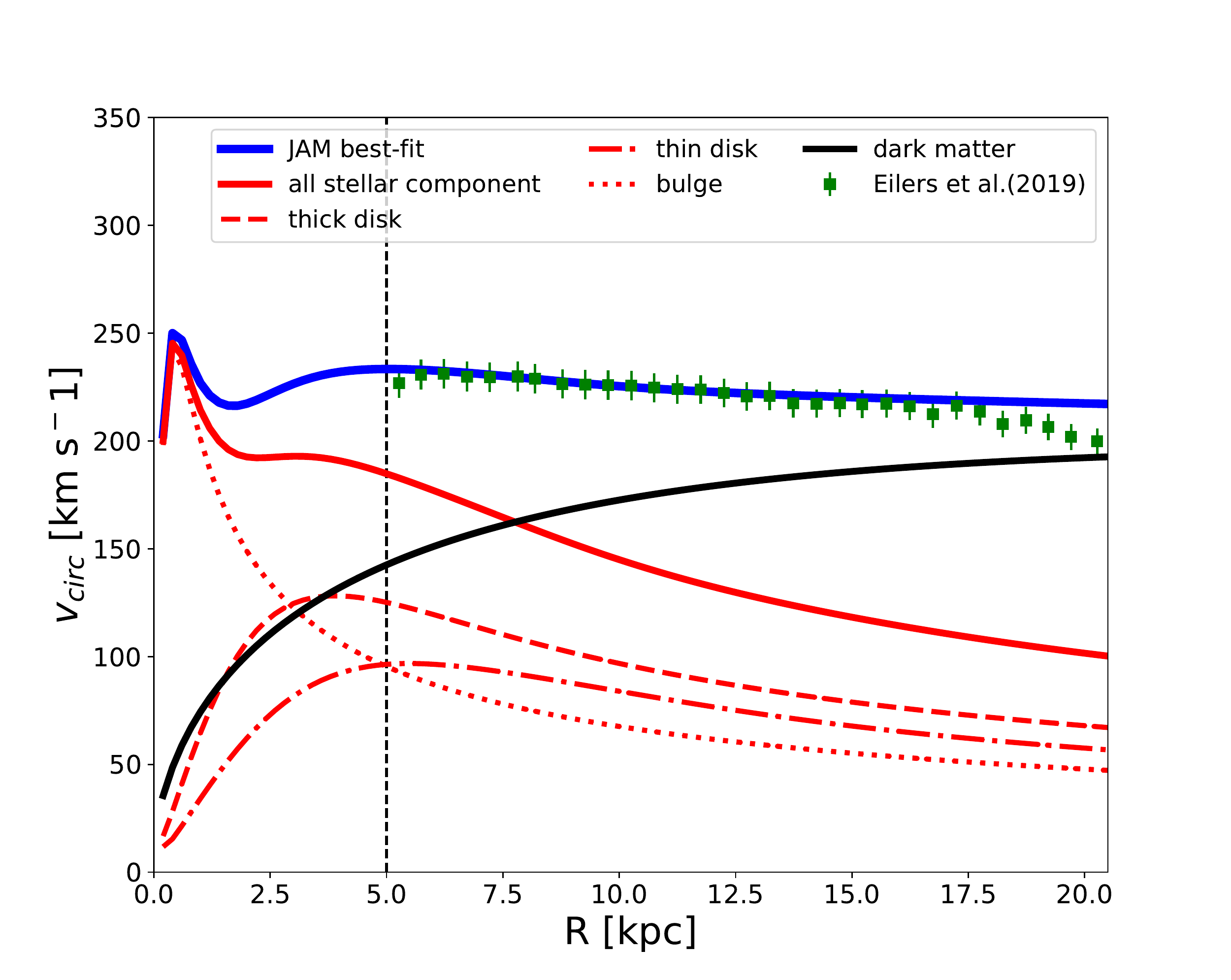}
 \caption{\textbf{Circular velocity curve same assumptions as in \citet{eilers18}.} The green dots are the measurements from \citet{eilers18} with 3 per cent systematic errors. Our circular velocity curve using the potential we get using JAM is shown in blue, the stellar component is shown as red solid line while the thin disk is a red dashed/doted line, the thick disk is a dashed line and the bulge is doted. The dark matter halo is shown as the black solid line. The inner part of the galaxy, i.e. $R<5$~kpc (indicated by the black dashed line), was excluded from the analysis by \citet{eilers18} due to the non-axisyemmtric influence of the Galactic bar.}
\label{im:circ eilers}
\end{figure}

Using the JAM$_{\mathrm{cyl}}$ \citep{JAM} and the above assumptions we can reproduce almost perfectly the velocity curve by \citet{eilers18}, see Figure~\ref{im:circ eilers} which is plotted as fig. 3 in \citet{eilers18} for comparison.

This, proves that the offset we have in our main model, is not caused by the modelling method but by the data set and the Galaxy model we assume. Small discrepancies are towards the Center where we do not have data and towards higher radii. The last one confirms our suspicion that the \citet{pouliasis17} model overestimates the mass in the inner parts and therefore underestimates the stellar mass towards larger radii to compensate it, which causes a high dark matter mass at larger radii.

\newpage
\thispagestyle{empty}
\mbox{}
\newpage
\bibliography{references}{}
\bibliographystyle{aasjournal}

\end{document}